\documentclass[%
aps,
 reprint,
 amsmath,amssymb,
prper,
floatfix
]{revtex4-2}

\usepackage{graphicx}% Include figure files
\usepackage{dcolumn}% Align table columns on decimal point
\usepackage{bm}% bold math
\usepackage{tikz}
\usepackage{pgfplots}
\usepackage[export]{adjustbox}
\pgfplotsset{compat=1.17}

\begin{document}

%\preprint{APS/123-QED}

\title{Grading Assistance for a Handwritten Thermodynamics Exam using Artificial Intelligence: An Exploratory Study}% Force line breaks with \\
%\thanks{A footnote to the article title}%

\author{Gerd Kortemeyer}
 \email{kgerd@ethz.ch}
 \affiliation{%
Rectorate and AI Center, ETH Zurich, 8092 Zurich, Switzerland
}%
\altaffiliation[also at ]{Michigan State University, East Lansing, MI 48823, USA}

\author{Julian N{\"o}hl}
 \email{jnoehl@ethz.ch}
 \affiliation{%
Institute of Energy \& Process Engineering, ETH Zurich, Tannenstrasse 3, 8092 Zurich, Switzerland
}%

\author{Daria Onishchuk}
 \email{donishchuk@ethz.ch}
 \affiliation{%
Unit for Teaching and Learning, ETH Zurich, 8092 Zurich, Switzerland
}%

\date{\today}% It is always \today, today,
             %  but any date may be explicitly specified

\begin{abstract}
Using a high-stakes thermodynamics exam as sample (252~students, four multipart problems), we investigate the viability of four workflows for AI-assisted grading of handwritten student solutions. We find that the greatest challenge lies in converting handwritten answers into a machine-readable format. The granularity of grading criteria also influences grading performance: employing a fine-grained rubric for entire problems often leads to bookkeeping errors and grading failures, while grading problems in parts is more reliable but tends to miss nuances. We also found that grading hand-drawn graphics, such as process diagrams, is less reliable than mathematical derivations due to the difficulty in differentiating essential details from extraneous information. Although the system is precise in identifying exams that meet passing criteria, exams with failing grades still require human grading. We conclude with recommendations to overcome some of the encountered challenges.
\end{abstract}

\maketitle

\section{Introduction}
\subsection{AI-supported Grading}
The recent advancements in artificial intelligence (AI) have ushered in new possibilities in various domains, including education. At its public appearance in fall 2022, Generative Pre-trained Transformer (GPT)~\cite{chatgpt} already demonstrated remarkable capabilities. Beyond the hype-wave triggered by the human appearance of its responses~\cite{turing1950}, it proved proficiency in academic domains such as physics, where it and later releases (in particular GPT-4~\cite{gpt4}) passed standardized exams and introductory courses at impressive levels~\cite{kung2022,achiam2023gpt,lawexam,kortemeyer23ai,polverini24,kortemeyer24cheating}. AI starts to get integrated into physics education~\cite{yeadon2024impact,sperling2024artificial}, where in addition to solving problems, it for example shows promise in constructing new physics problems~\cite{sperling2024artificial,kuchemann23}, making physics materials accessible to blind readers~\cite{kortemeyer2023using}, and supporting physics educational research efforts~\cite{tschisgale2023integrating,kieser23}.

The grading of physics exams has traditionally required deep analytical skills to assess not just the final answers but the problem-solving process itself, a task that involves evaluating logical, conceptual, and mathematical competencies~\cite{reif1976,reif1995,hsu2004,hattie2008,alsalmani23}. While certain aspects of problem solutions can be graded by computers, a thorough evaluation often necessitates human judgment, particularly when diverse solution paths and potential errors are involved~\cite{kashyd01,kortemeyer08,risley2001,stelzer2001,dufresne02,fredericks2007,richards2011,perdian2013,docktor2016,burkholder2020}. Such exams usually require handwriting solution paths, since in exam situations, typesetting those equations would take up an inordinate amount of time for the students, and it would necessitate a skillset outside the learning objectives of most physics courses.

\begin{figure}
\begin{center}
\includegraphics[width=\columnwidth]{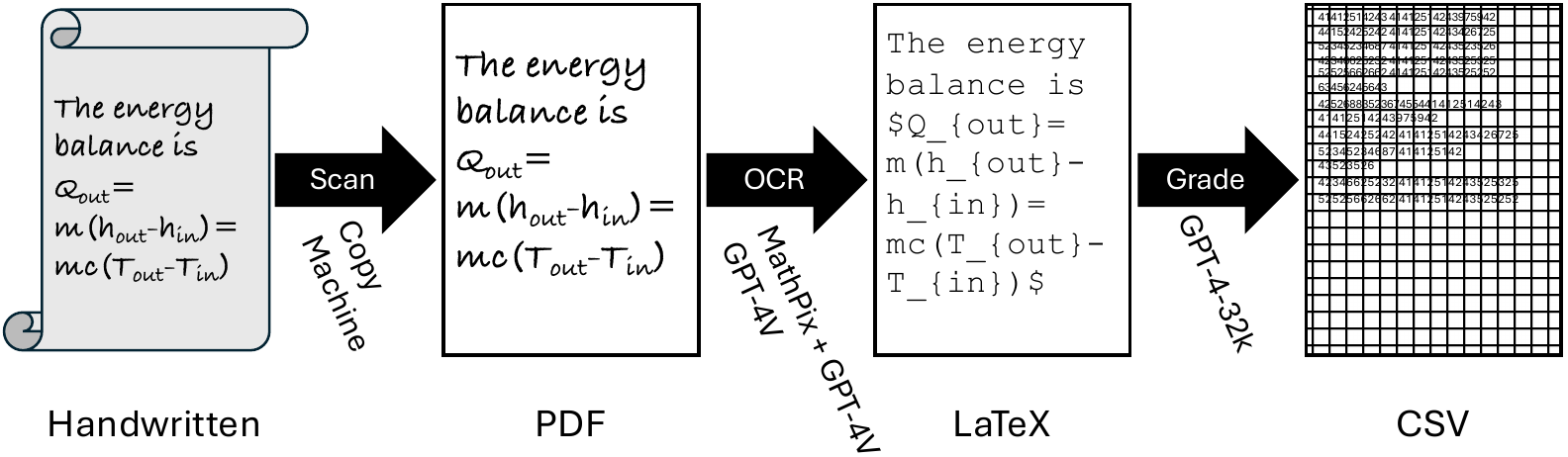}
\end{center}
\caption{Workflow of exam grading, starting with scanning the exam sheets, making them machine-readable with MathPix~\cite{mathpix} and GPT-4V~\cite{gpt4}, and grading them using GPT-4-32k.}
\label{fig:workflow}
\end{figure}

AI holds potential as a scalable solution for giving feedback on open-ended responses~\cite{wan24}, as well as grading, or at least classification~\cite{wilson22} and pre-grading, by leveraging its ability to process large volumes of data and its emerging capability to understand and evaluate complex student responses~\cite{mitros2013,dzikovska2013,burrows2015,sung2019,azad2020,fowler2021,stanyon2022,grassini2023,kortemeyer24aigrading}. In a recent study using a synthetic data set, an agreement of $R^2=0.84$ could be found between human and AI grades~\cite{kortemeyer24aigrading}, however, the synthetic nature of the investigated solution derivations circumvented many of the intricacies that are present in authentic exam situations.  Most notably, challenges are introduced by the complexity of handwritten mathematical expressions, which add another layer of difficulty due to the intricacies of Optical Character Recognition (OCR) technology~\cite{mori1992historical,okamura1999handwriting,wang2021}. Figure~\ref{fig:workflow} shows the workflow that we investigated, using a large-scale high-stakes physics exam. We focus on the operational challenges involved, as well as the limitations and capabilities of current technology to augment traditional grading methods.

\subsection{Large Language Models and Randomness}
Large Language Models (LLMs) like those based on the architecture introduced by Vaswani et al.~\cite{vaswani2017attention} use probabilistic methods to generate responses. They create sequences of  so-called tokens (in the English language similar to syllables).  Even responses to clear-cut physics-related prompts can vary strongly~\cite{kortemeyer23ai,kortemeyer24aigrading}, influenced by a temperature parameter $T$. A lower $T$ results in more predictable outputs, while a higher $T$ increases creativity. At absolute zero ($T=0$), a language model consistently selects the most probable next token, making its outputs highly predictable though not necessarily accurate due to potential biases or errors in the training data~\cite{vaswani2017attention,renze2024effect}. Conversely, a higher $T$ can lead to less reliable, ``hallucinated'' responses (``hallucination'' in this context refers to a phenomenon where the model generates output that is not grounded in the input data or reality). In this study, the default setting was $T=0.7$, except for OCR of complete pages, where it was set to $T=0.5$. Our study assumes that grading by LLMs should be viewed probabilistically. Multiple runs yield average scores, with standard deviations providing a measure of confidence. For instance, a score of $14 \pm 5$ suggests lower reliability than $14 \pm 2$, although the score could still be incorrect.
 
\section{Setting}
\subsection{Institution}
ETH Zurich is a technical university with approximately 25,000 students from 120 countries, with about one-third identifying as female. Admission is highly selective for international students, yet unrestricted for anyone holding a Swiss high school diploma. Most undergraduate courses are taught in German. In the German-speaking universities' academic tradition, ETH Zurich emphasizes summative assessments at the end of courses, rather than smaller assessments throughout the course. The study was  approved by the  institution's Ethics Committee (IRB) under 2023-N-286.

\subsection{Cloud Infrastructure}
Access to OpenAI models was provided through Azure AI Services~\cite{azure}, where ETH Zurich has a contract which assures processing on Swiss data centers, with a consumption-based per-token payment structure. Based on our subscription, MathPix was likely processing in the United States, which is why we used pseudonyms.

\subsection{Exam}
We considered a high-stakes exam for engineers on thermodynamics, dealing with standard topics of energy, exergy, entropy, and enthalpy. Students had 15~minutes to prepare by reading the problems and two hours to complete the work. They could use provided reference materials and a non-programmable calculator. They needed to provide handwritten solutions including derivations, using permanent pens and scribbling out anything they did not want graded. A total of 252 out of 434 students taking the exam agreed to participate in the study.

Problem~1 involves a reactor in steady-state operation, including liquid in- and outflow, a chemical reaction with associated heat generation, and a cooling jacket. The problem parts are:
\begin{enumerate}
    \item[a)] Calculation of the heat transfer to the cooling fluid (2.5 points).
    \item[b)] Determination of the thermodynamic mean temperature of the cooling fluid (2.5 points).
    \item[c)] Calculation of the entropy production due to heat transfer (1.5 points).
    \item[d)] After steady-state operation, the outlet flow is stopped.  Calculation of the water mass required for cooling the reactor from its operating to a lower temperature using an energy balance (4.5 points).
    \item[e)] Determination of the change in entropy of the reactor contents between the initial and cooled states (4 points).
\end{enumerate}

Problem~2 explores the operation of an aircraft engine, consisting of reversible or irreversible compressors, turbines, nozzles and heat addition. The problem parts are:
\begin{enumerate}
    \item[a)] Drawing of the engine process qualitatively in a T-s diagram, marking relevant states (8 points).
    \item[b)] Determination of the exit speed and temperature of the aircraft engine (4 points).
    \item[c)] Calculation of the specific exergy increase between two states (3.5 points).
    \item[d)] Calculation of the specific exergy loss related to the mass flow rate (3.5 points).
\end{enumerate}

Problem~3 involves a hot gas and a solid-liquid system in an isolated cylinder separated by a heat-transferring membrane. The problem parts are:
\begin{enumerate}
    \item[a)] Calculation of the initial pressure and mass of the gas in the cylinder  (5.5 points).
    \item[b)] Determination of the temperature and pressure of the gas after an equilibrium state has been reached through heat transfer (2 points).
    \item[c)] Calculation of the heat transferred from the gas to the ice-water mixture (2.5 points).
    \item[d)] Calculation of the ice content in the second state (7 points).
\end{enumerate}

Problem~4 describes a two-step freeze-drying process for food preservation. The problem parts are:
\begin{enumerate}
    \item[a)] Drawing of the freeze-drying process in a p-T diagram, including labeled phase regions (4.5 points). 
    \item[b)] Determination of the required mass flow rate of the refrigerant R 134a (6 points).
    \item[c)] Determination of the vapor fraction of the refrigerant immediately after throttling (4 points).
    \item[d)] Calculation of  the coefficient of performance for the cooling cycle (3 points).
    \item[e)] Discuss how the temperature inside the freeze-dryer changes if the cooling cycle continues unchanged (1 point).
\end{enumerate}

\begin{figure*}
\begin{flushright}
\includegraphics[width=0.46\textwidth]{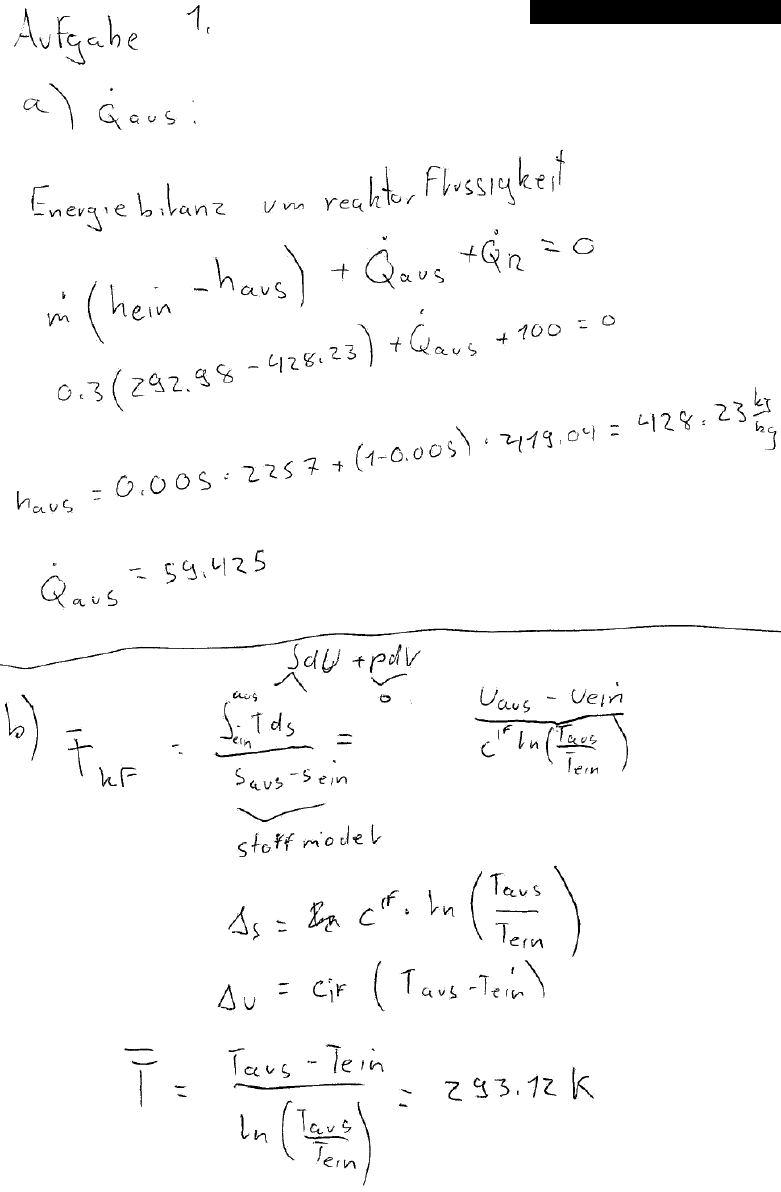}\qquad\raisebox{4cm}[0pt][0pt]{\includegraphics[width=0.46\textwidth]{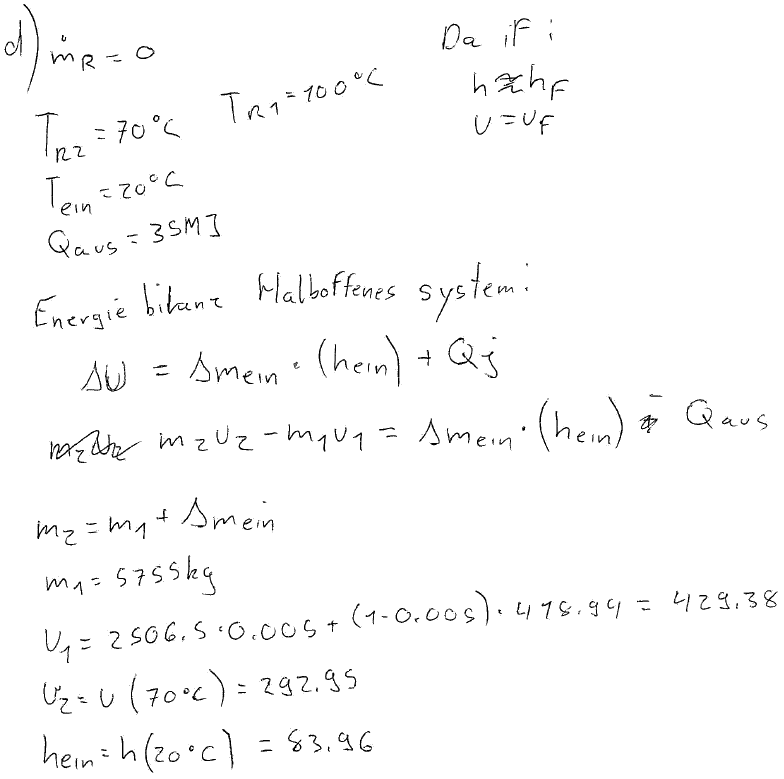}}

\includegraphics[width=0.46\textwidth]{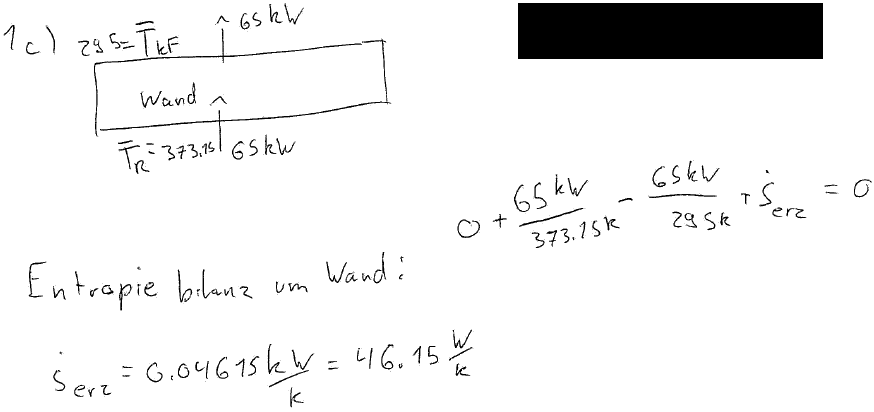}\qquad\raisebox{1cm}[0pt][0pt]{\includegraphics[width=0.46\textwidth]{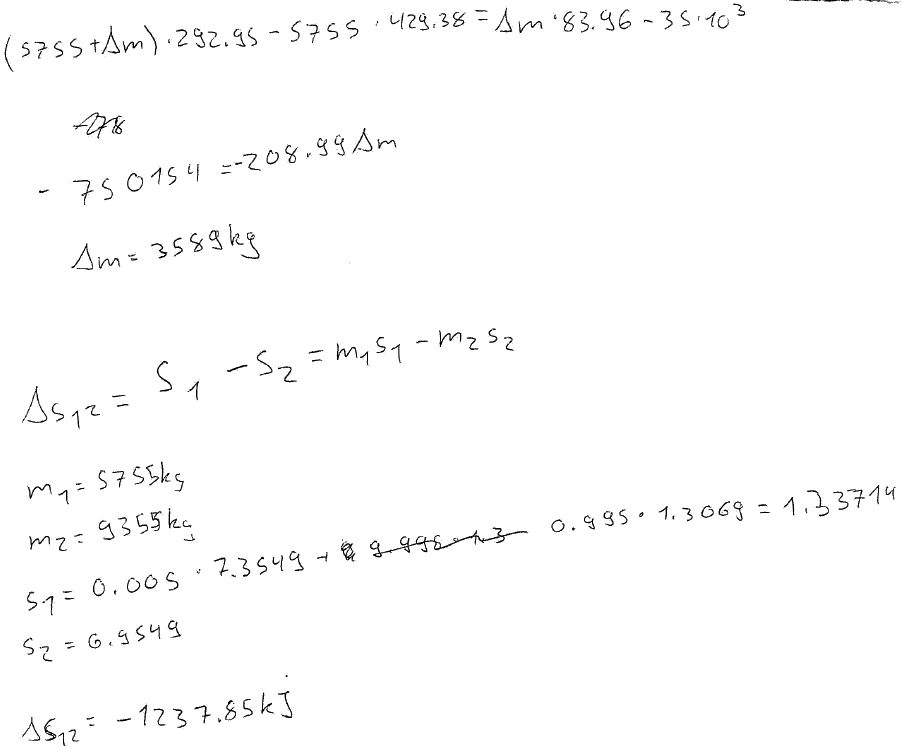}}
\end{flushright}
\caption{Example of a handwritten solution for Problem~1. The handwriting itself is fairly clear, but as students were instructed to simply scribble out errors, this solution, like many others, contains several scratched out symbols and expressions.}
\label{fig:prob1hand}
\end{figure*}

\begin{figure}
\begin{center}
\includegraphics[width=\columnwidth]{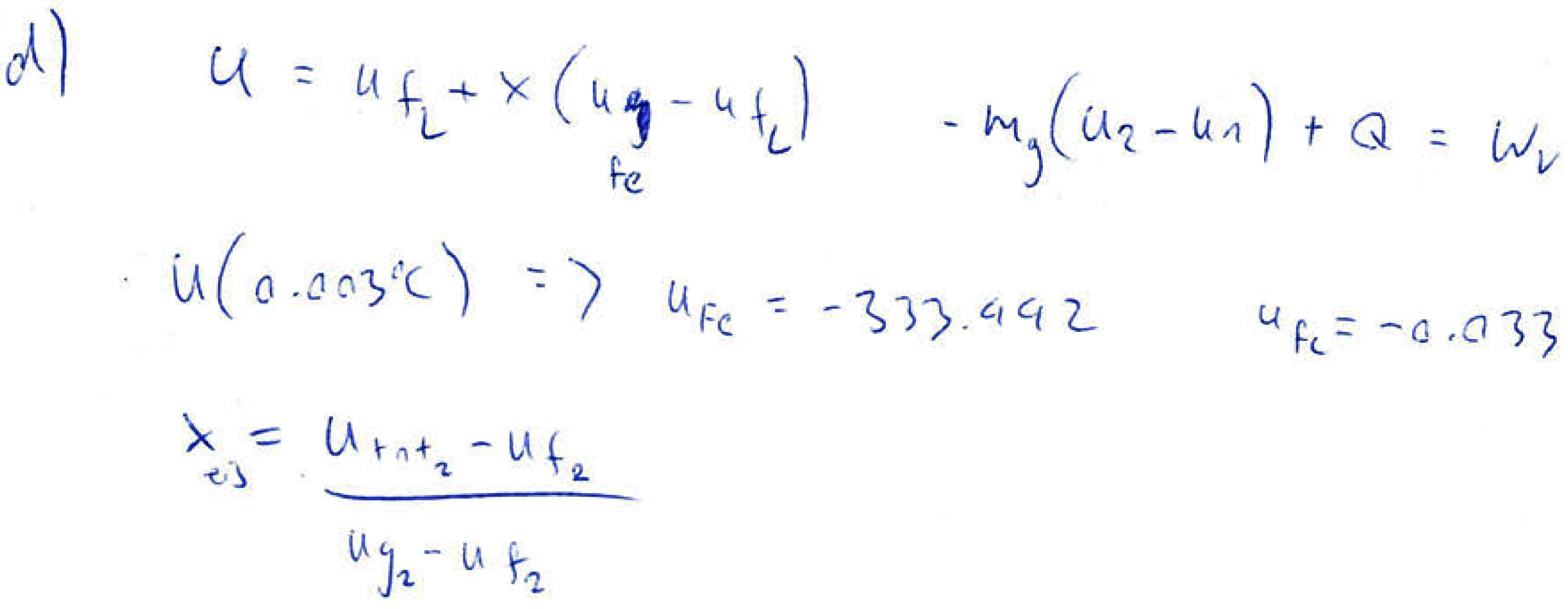}
\end{center}
\caption{Example of a handwritten solution for Problem~3d. Arguably, such unfortunately very typical handwriting is hard to decipher without knowing the  context.}
\label{fig:examplehand}
\end{figure}

Figure~\ref{fig:prob1hand} shows a typical example for a handwritten solution to Problem~1 (the black box on top resulted from redacting the student name). Where the student noted ``TAB A-2,'' he or she referred to the data table that was available during the exam. Figure~\ref{fig:examplehand} shows another, less well readable example of a solution for Problem~3d.

A feature of the German-tradition university system is that exams can be passed or failed (though, at ETH Zurich, for some exams this gets decided on the level of so-called exam blocks with compensation mechanisms), and a notion among some students in the beginning semesters is that they do not really care about the grade as long as they pass the exam and can move on with their studies. Due to the high-stakes nature of these exams, every exam problem is graded by two teaching assistants. 

The exam was graded on a fine-grained rubric by the TAs, awarding points for each solution step. The course personnel creating and correcting each task defines sub-points which a student should at least get to pass the exam. These points should show a basic understanding of Thermodynamics. The sum of the points is used to define the lower boundary for the ``passing grade'' of 4.0.  The value for the highest grade of 6.0 is based on values from previous exams (\% of total points). However, slight adjustment to the mapping of points to grades can be made to coincide with gaps in the point
distribution. As a result, the nominal level of passing the exam was set at 26~out of the~65 available points (since there are compensation mechanisms, we use the term ``nominal''). If the average grades of a student are below a certain threshold so that he or she needs to repeat the semester, the exam was regraded by a third teaching assistant.

\section{Methodology}
\subsection{Workflow and Study-specific Additional Steps}
Two important directives for this study were non-interference with the regular exam process and preservation of student privacy. This meant that students were not given any instructions on how to prepare their work for better OCR interpretation and AI-grading, which brought about additional challenges and problems that had to be fixed manually. Grading personnel was not informed which students had given informed consent to participate in the study. After the exam packages had been turned in, the investigators separated the consent forms from the exam sheets and scanned only the solutions of the students who had agreed to participate in the study.  The scanner device had automatic paper handling capabilities (as it turned out, it also automatically switched between black-and-white mode (see Fig.~\ref{fig:prob1hand} as an example) and color-mode (see Fig.~\ref{fig:examplehand}), and it automatically skipped empty pages); the device then emailed the PDFs to the investigators.

The workflow included a pseudonymization step that would not be present in production use. The investigators and the grading personnel initially shared a key, where each exam package had a number. One of the investigators (GK) redacted all names from the scanned solutions and inserted the exam-package number before any further processing. As exam sheets within the package were often out-of-order, at this step, the PDF-pages were manually sorted, and clear-text markers were inserted to separate the problems. The latter was hampered by students frequently only writing letters like ``c)'' at the start of a new problem part, but not the problem number; if the sheets were out of order and had no page numbers written, this meant figuring out from the contents if this was Part~c of Problem~1,~2,~3, or~4.

Students had brought their own paper to the exam, which was frequently repurposed and included company letterheads with logos or advertising or had pre-printed headers like ``My Notes'' or ``While you were out.'' While these page augmentations would not interfere with the TA-grading, they had to be redacted by hand, as they would otherwise be transcribed by the OCR process and appear in the middle of solution derivations. Some students also used their paper in a multi-column landscape layout; these pages appeared portrait-oriented with~90 or 270~degree rotated writing, and had to be rotated into landscape format by hand. Other students turned their two-sided sheets along the short instead of the long side when continuing work on a problem, which resulted in every other page being upside down.

During the OCR step, the exam-package number and the problem-separation markers were transcribed into the LaTeX, since they simply appeared as plain text. Subsequent scripts would pick up those markers from the LaTeX for processing. Grading personnel eventually reported their grading decisions to the investigators in terms of the exam-package number for comparison.

\subsection{Optical Character Recognition (OCR)}
Figure~\ref{fig:prob1hand} shows an example for a fairly readable solution, while Fig.~\ref{fig:examplehand} provides an example for less clear handwriting.
To render these solutions machine-readable for LaTeX processing, two approaches were investigated, using scripts and the APIs of the respective systems:

\subsubsection{MathPix plus GPT-4V}
\begin{figure}[!ht]
\begin{center}
\includegraphics[width=0.64\columnwidth]{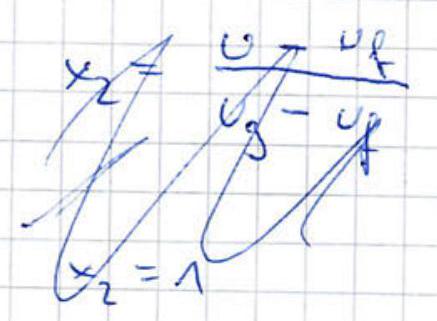}
\end{center}
\caption{Example of a cropped JPG produced by MathPix from a student solution of Problem~3d. }
\label{fig:oddity}
\end{figure}

The first method employed MathPix~\cite{mathpix} for the preliminary conversion of handwritten content into LaTeX. MathPix performs a robust and rather deterministic interpretation of handwriting and mathematical expressions, and it does not hallucinate, but it is also designed not to guess when confidence is low. Instead of producing textual output, for segments of a document where any interpretation would be  low-confidence, or in situations where graphs, figures, or sketches were identified, MathPix incorporates a cropped JPG image into the LaTeX document via the \texttt{\textbackslash includegraphics} command.  While usable in many other scenarios, the grading step needs a fully textual representation, where at least some interpretation of unclear handwriting is provided, and where graphical content is described in text (the latter is crucial for Problem~2a and Problem~4a, where diagrams are expected). Thus,  subsequently, using another script, these JPGs were processed by multimodal GPT-4V~\cite{gpt4}, which generated LaTeX code to replace the \texttt{\textbackslash includegraphics} commands. Quite often, though, these image identified by MathPix turned out to be scribbled-out calculations, like the one in Fig.~\ref{fig:oddity};  while unrecognizable to MathPix, scribbled-out content frequently cannot be distinguished from drawings, annotations, scanning noise, paper markings, or simply unclear writing by GPT-4V. As a result, these images were sometimes read and inserted as if the scribbling-out did not exist (thus including content into the grading that the student had discarded), and sometimes completely misinterpreted.

\begin{figure}
\begin{center}
\includegraphics[width=0.8\columnwidth]{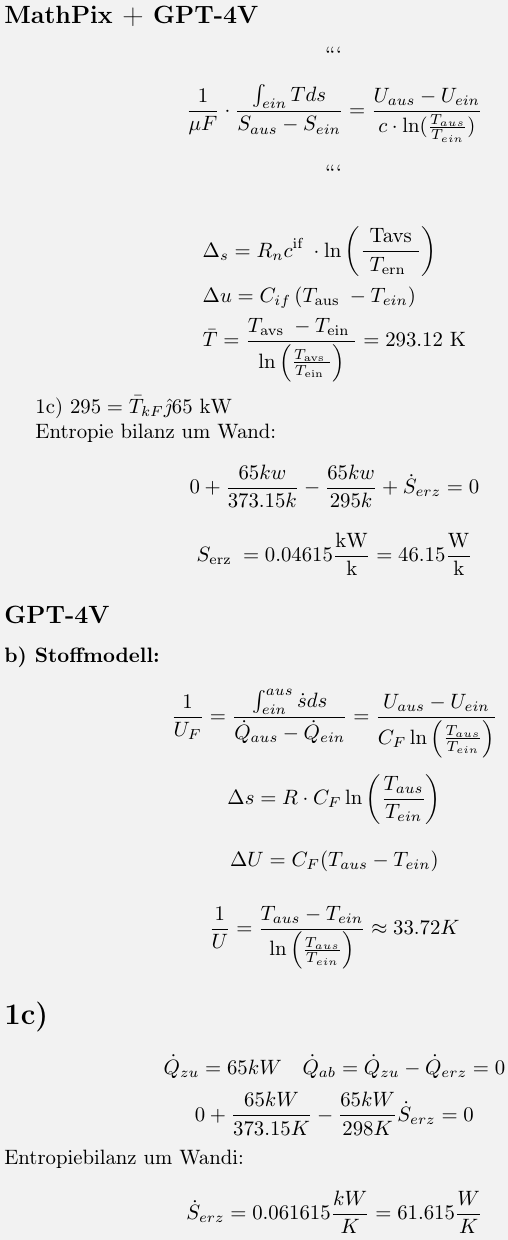}
\end{center}
\caption{Different OCR renderings of the student solution for Problem~1b and~c  in Fig.~\ref{fig:prob1hand}.}
\label{fig:ocr2}
\end{figure}

\begin{figure}
\begin{center}
\includegraphics[width=0.9\columnwidth]{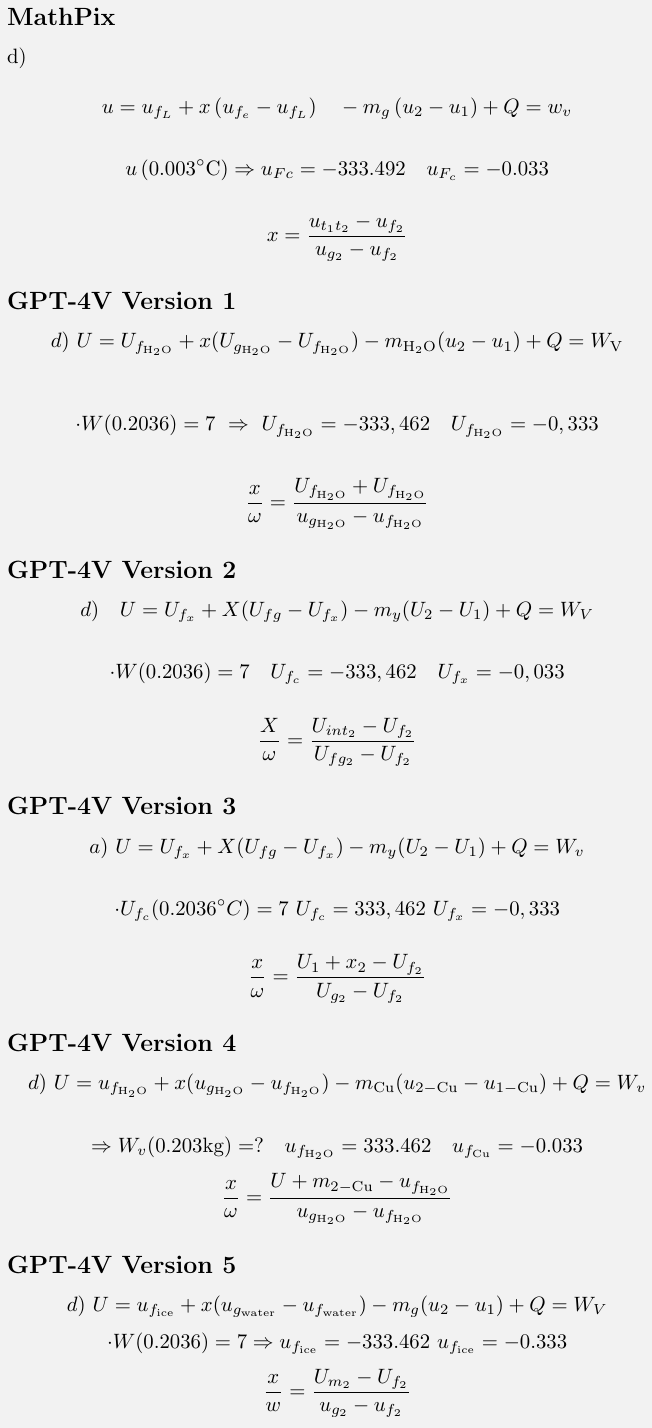}
\end{center}
\caption{Different OCR renderings of the student solution for Problem~3d  in Fig.~\ref{fig:examplehand} ($T=0.5$).}
\label{fig:ocr}
\end{figure}

The top panel of Figure~\ref{fig:ocr2} shows an example of the whole process for parts~b and~c of Problem~1, shown in Fig.~\ref{fig:prob1hand}. MathPix was unable to process the first expression in Part~b, likely because of the annotations above and below the formula terms. The expression was thus turned into an image and subsequently processed by GPT-4V. MathPix, however, did not recognize the small drawing in Part~c as such and thus itself processed the handwriting. The top panel of Fig.~\ref{fig:ocr} shows MathPix's interpretation of the solution in Fig.~\ref{fig:examplehand}, which is accurate on the level that a human would decipher the handwriting without physics knowledge what to expect. The subscript of $x$, ``eis,'' is the German word for ``ice,'' which would have been nearly impossible to figure out without knowing what the problem is about. The readability is further hampered by the use of a ball pen, which did not provide continuous lines.

The process was slightly hampered by Azure at various times declining to process image descriptions, for example the image shown in Fig.~\ref{fig:oddity} produced the error, ``the response was filtered due to the prompt triggering Azure OpenAI's content management policy'' for whatever GPT-4V saw in the image. 

\begin{figure}[!ht]
\noindent\fbox{
\begin{minipage}{\columnwidth}
\includegraphics[width=\linewidth]{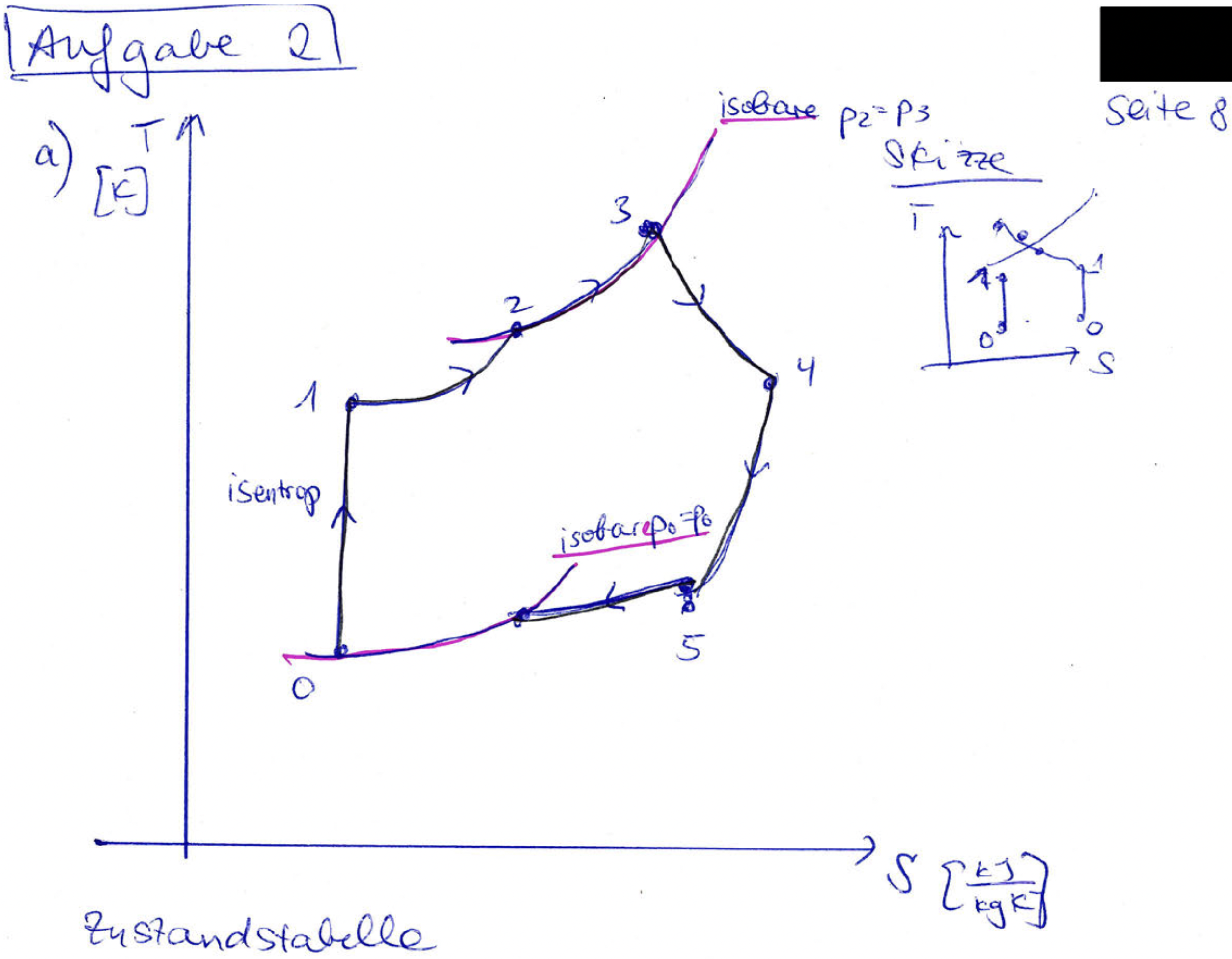}
\begin{flushleft}
\footnotesize
The image shows a graphical representation, likely some kind of diagram. The horizontal axis is marked with an "S", which might stand for distance or another variable. The vertical axis is labeled with an "I", which could represent intensity or a similar quantity. A curve is drawn across the axes, marked at three points, probably to highlight specific values or points. Between these are two vertical arrows, one labeled with a "1"; the other is smaller and marked with a "0". Additionally, there is a horizontal arrow pointing from right to left. Above it, there is a curve ascending from left to right with an arrow indicating the direction of the curve.
\end{flushleft}
\end{minipage}
}
\caption{GPT-4V-generated description (English translation of original German) of a student solution for Problem~2a.}
\label{fig:studentgraphdesc}
\end{figure}

Figure~\ref{fig:studentgraphdesc} shows an example of a description for a graph drawn by a student in response to Problem~2a, generated by GPT-4V based on an image identified by MathPix. Unfortunately, quite often, MathPix ignored these graphs if they came too close to the margin of the page or where horizontally aligned with mathematical expressions.

\subsubsection{Using only GPT-4V}
The alternative method involved converting entire PDF pages to PNG format and processing these solely with GPT-4V; to account for the variability, this was done several times. The bottom panel of Fig.~\ref{fig:ocr2} shows an example of this process for parts~b and~c in Fig.~\ref{fig:prob1hand}. Since due to token-limits, GPT-4V  could only process one page image at a time, this entailed uploading each individual page multiple times for interpretation and then reassembling the LaTeX outputs. We decided to assemble all first interpretations of individual pages into a first version of the whole exam, all second interpretations into a second version, etc.; however, this was arbitrary, since all interpretations were independent and, for example,  the third interpretation of the fourth page would have no particular connection to the third interpretation of the fifth page. 

This approach exhibited strong variability in the accuracy of text interpretation, occasionally producing erroneous interpretations and extrapolations. To limit the variations, the temperature was lowered to $T=0.5$, but there were still strong variations, for example:
\begin{itemize}
\item As the bottom panel of Fig.~\ref{fig:ocr2} shows, all final result numbers were misread.
\item The lower panels of Fig.~\ref{fig:ocr} show five interpretations of the handwriting in Fig.~\ref{fig:examplehand}.  GPT-4V shows remarkable creativity interpreting the unclearly written subscripts; MathPix completely skipped ``eis,'' which overall may be the safer choice.
\end{itemize}

The temperature of $T=0.5$ was determined based on a small sample size; obviously, finding the optimum temperature would be the topic of further investigations, but computationally intense.
Unfortunately, it turned out graphs like the one in Fig.~\ref{fig:studentgraphdesc} were also ``overlooked'' by GPT-4V when processing a complete page, making the graph transcription process rather unreliable.

Overall, it was found that MathPix with GPT-4V support was more reliable and reproducible  than using only GPT-4V. Since in addition, processing whole pages with GPT-4V turned out to be very token-intensive, and with GPT-4V tokens being more expensive than regular GPT-4-32k tokens, it was decided to only process a subset of 20~exams, but do so five times to account for the strong variability in the results.

\subsection{Grading Step}
Following the conversions, the generated LaTeX documents were graded by GPT-4-32k. In addition, the system was prompted to provide a justification for its decisions. Several approaches were evaluated using the first problem to gauge their viability. For two of the more promising approaches, the complete exam was graded. For all grading approaches, the LLM was provided with the problem text itself, but the grading approaches differed by the granularity of their grading scope and sample solution documents.

\subsubsection{Rubric-based Grading}
The teaching assistants (TAs) used a fine-grained grading rubric, which operated on the level of points and half-points. Fig.~\ref{fig:sample} shows an example of such a grading rubric, where the partial points are marked in red. For AI-grading, the derivation steps in the sample solution were marked up with point values and criteria, and this list was provided to the LLM as a table (see Table~\ref{tab:grading_rubric} as an example). While the TAs were supposed to make binary decisions whether or not to provide the rubric points, it was realized very early on that the LLM functioned less arbitrarily on a sliding scale, allowing points like 0.4.
\begin{table*}
\caption{Example of the grading rubric for the first exam problem.}
\begin{tabular}{lllp{13cm}}
Part&Item&Max.&Criterion\\
&&Pts.&\\
\hline
a&	a\_Bilanz&	1&	Maximum points (1.0) for correct calculation of heat flow to the cooling jacket; energy balance around the reactor. Full points for accurately applying \(0=\dot{m}_{\mathrm{ein}} \cdot h_{\mathrm{ein}}-\dot{m}_{\mathrm{aus}} \cdot h_{\mathrm{aus}}+\dot{Q}_{\mathrm{R}}-\dot{Q}_{\mathrm{aus}}\) with \(\dot{m}_{\text {ein}}=\dot{m}_{\text {aus}}=\dot{m}\). Less points for correct approach with minor errors.\\
a&	a\_h\_ein&	0.5&	Maximum points (0.5) for correct calculation of \(h_{\text {ein}}=h_{\mathrm{f}}(70^{\circ} \mathrm{C})=292.98 \frac{\mathrm{kJ}}{\mathrm{kg}}\). Less points for minor numerical errors.\\
a&	a\_h\_aus&	0.5&	Maximum points (0.5) for correct calculation of \(h_{\text {aus}}=h_{\mathrm{f}}(100^{\circ} \mathrm{C})=419.04 \frac{\mathrm{kJ}}{\mathrm{kg}}\). Less points for minor numerical errors.\\
a&	a\_Erg&	0.5&	Maximum points (0.5) for correct calculation of \(\dot{Q}_{\text {aus}}=0.3 \frac{\mathrm{kg}}{\mathrm{s}} \cdot (292.98 \frac{\mathrm{kJ}}{\mathrm{kg}}-419.04 \frac{\mathrm{kJ}}{\mathrm{kg}})+100 \mathrm{~kW}\). Less points for minor errors.\\
b&	b\_T\_KF\_Tds&	1&	Maximum points (1.0) for correct derivation of the thermodynamic mean temperature of the cooling water flow, \( \bar{T}_{\mathrm{KF}} \). Full points for correct formula and calculation. Less points for minor errors.\\
b&	b\_Stoffmod&	1&	Maximum points (1.0) for correct application of the ``ideal fluid'' model. Full points for correct use of either equation provided. Less points for minor calculation errors.\\
b&	b\_Erg&	0.5&	Maximum points (0.5) for correct calculation of \( \bar{T}_{\mathrm{KF}} =\frac{T_{\text {aus }}-T_{\text {ein }}}{\ln (\frac{T_{\text {aus }}}{T_{\text {ein }}})}\). Less points for minor errors.\\
c&	c\_Bilanz&	1&	Maximum points (1.0) for correct entropy production calculation in heat transfer between reactor and cooling jacket. Full points for accurate balance equation application. Less points for minor errors.\\
c&	c\_Erg&	0.5&	Maximum points (0.5) for correct calculation of \( \dot{S}_{\text {erz }} =\frac{\dot{Q}_{\text {aus }}}{\bar{T}_{\mathrm{KF}}}-\frac{\dot{Q}_{\text {aus }}}{T_{\text {Reaktor }}}\). Less points for minor numerical errors.\\
d&	d\_u\_f&	0.5&	Maximum points (0.5) for correct calculation of \( u_{\mathrm{f}}(100^{\circ} \mathrm{C}) = 418.94 \frac{\mathrm{kJ}}{\mathrm{kg}} \). Less points for numerical inaccuracies.\\
d&	d\_u\_g&	0.5&	Maximum points (0.5) for correct calculation of \( u_{\mathrm{g}}(100^{\circ} \mathrm{C}) = 2506.5 \frac{\mathrm{kJ}}{\mathrm{kg}} \). Less points for numerical inaccuracies.\\
d&	d\_u\_1&	1&	Maximum points (1.0) for correct calculation of \( u_{1} = x_{D} \cdot u_{\mathrm{g}}(100^{\circ} \mathrm{C})+(1-x_{D}) \cdot u_{\mathrm{f}}(100^{\circ} \mathrm{C}) \). Less points for minor errors.\\
d&	d\_u\_2&	0.5&	Maximum points (0.5) for correct calculation of \( u_{2} = u_{\mathrm{f}}(70^{\circ} \mathrm{C}) = 292.95 \frac{\mathrm{kJ}}{\mathrm{kg}} \). Less points for numerical errors.\\
d&	d\_h\_ein&	0.5&	Maximum points (0.5) for correct calculation of \( h_{\text {ein }}=h_{\mathrm{f}}(20^{\circ} \mathrm{C})=83.96 \frac{\mathrm{kJ}}{\mathrm{kg}} \). Less points for minor numerical errors.\\
d&	d\_Bilanz&	1&	Maximum points (1.0) for correct energy balance in an open system for determining added water mass. Full points for accurate calculation. Less points for incorrect or partial calculations.\\
d&	d\_Erg&	0.5&	Maximum points (0.5) for correct calculation of change in mass \( \Delta m_{12} = 3756.84 \mathrm{~kg} \). Less points for minor errors.\\
e&	e\_s\_f&	0.5&	Maximum points (0.5) for correct entropy value of fluid at \( 100{ }^{\circ} \mathrm{C} = 1.3069 \frac{\mathrm{kJ}}{\mathrm{kg} \mathrm{K}} \). Less points for minor errors.\\
e&	e\_s\_g&	0.5&	Maximum points (0.5) for correct entropy value of gas at \( 100{ }^{\circ} \mathrm{C} = 7.3549 \frac{\mathrm{kJ}}{\mathrm{kg} \mathrm{K}} \). Less points for minor errors.\\
e&	e\_s\_1&	1&	Maximum points (1.0) for correct calculation of \( s_{1} = x_{D} \cdot u_{\mathrm{g}}(100^{\circ} \mathrm{C})+(1-x_{D}) \cdot u_{\mathrm{f}}(100^{\circ} \mathrm{C}) \). Less points for minor errors.\\
e&	e\_s\_2&	0.5&	Maximum points (0.5) for correct calculation of \( s_{2} = s_{\mathrm{f}}(70^{\circ} \mathrm{C}) = 0.9549 \frac{\mathrm{kJ}}{\mathrm{kg} \mathrm{K}} \). Less points for minor errors.\\
e&	e\_Bilanz&	1&	Maximum points (1.0) for correct entropy balance calculation \( \Delta S_{12} = S_{2} - S_{1} \). Full points for correct application of the formula. Less points for minor errors.\\
e&	e\_Erg&	0.5&	Maximum points (0.5) for correct calculation of \( \Delta S_{12} = 1387.62 \frac{\mathrm{kJ}}{\mathrm{K}} \). Less points for minor errors.
\end{tabular}

\label{tab:grading_rubric}
\end{table*}

\begin{figure}[!ht]
\noindent\fbox{
\begin{minipage}{\columnwidth}
\includegraphics[width=0.8\linewidth]{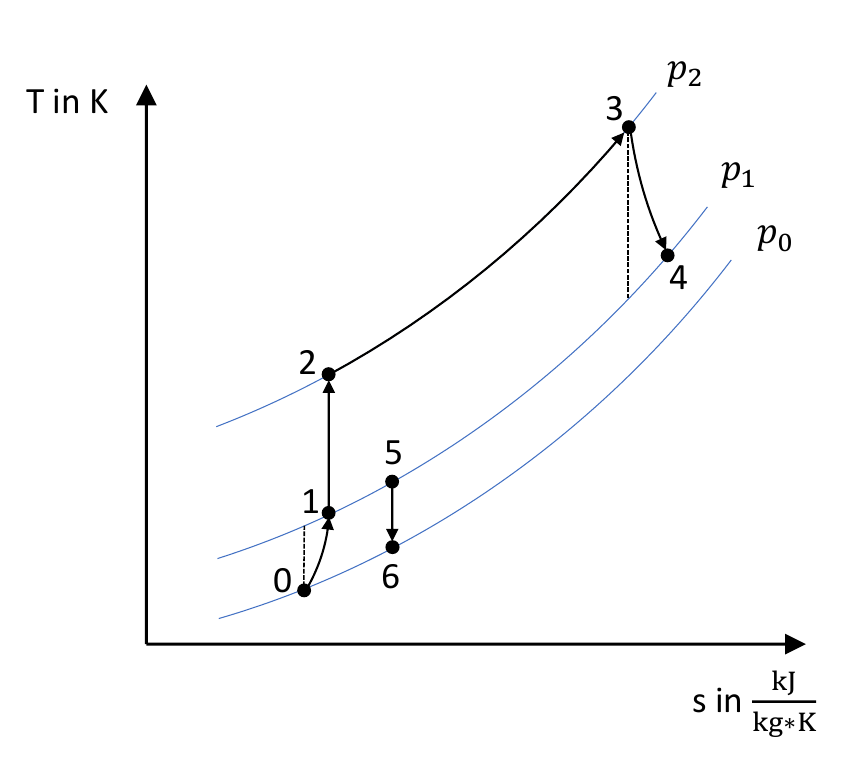}
\begin{flushleft}
\footnotesize
The diagram represents a thermodynamic cycle on a Temperature-Entropy ($T$-$s$) chart. The vertical axis is labeled as "$T$ in K" representing temperature in Kelvin, and the horizontal axis is labeled as "$s$ in $\text{kJ/kg}\cdot K$" representing specific entropy in kilojoules per kilogram-Kelvin. There are three isobars shown as curved lines, labeled from bottom to top as $p_0$, $p_1$, and $p_2$, indicating different pressure levels.

The cycle consists of six states, marked as points 0 to 6. Points are shown as solid black circles, and the processes between them are depicted as solid black lines with arrows indicating the direction of the process. Additionally, there are red circled numbers placed next to each process, which likely indicate the sequence of steps or the process number in a cycle. 

Starting from state 0, the process moves vertically up to state 2, indicating a constant entropy (isentropic) compression. From state 2 to 3, the process follows along the $p_2$ isobar, suggesting isobaric heat addition. The process from 3 to 4 is a vertical line downwards, showing an isentropic expansion. Then, the process from 4 to 5 follows the $p_0$ isobar, indicating isobaric heat removal. Finally, the process from 5 to 6 and 6 to 0 is shown as a two-step process with a vertical line down to state 6 and a horizontal line back to state 0, completing the cycle.
\end{flushleft}
\end{minipage}
}
\caption{GPT-4V-generated description of the process diagram in the sample solution of Problem~2a.}
\label{fig:graphdesc}
\end{figure}

\subsubsection{Parts-based Grading}
Here, the system was prompted with the sample solution instead of a fine-grained rubric, and only the total points for each part were given; for the first problem, this resulted in five point values for parts a--e instead of 22~rubric items. Overall, this approach turned out to be less computationally intense than rubric-grading, so it was applied to all problems, not just the first one.

\begin{figure}[!ht]
\begin{center}
\includegraphics[width=\columnwidth]{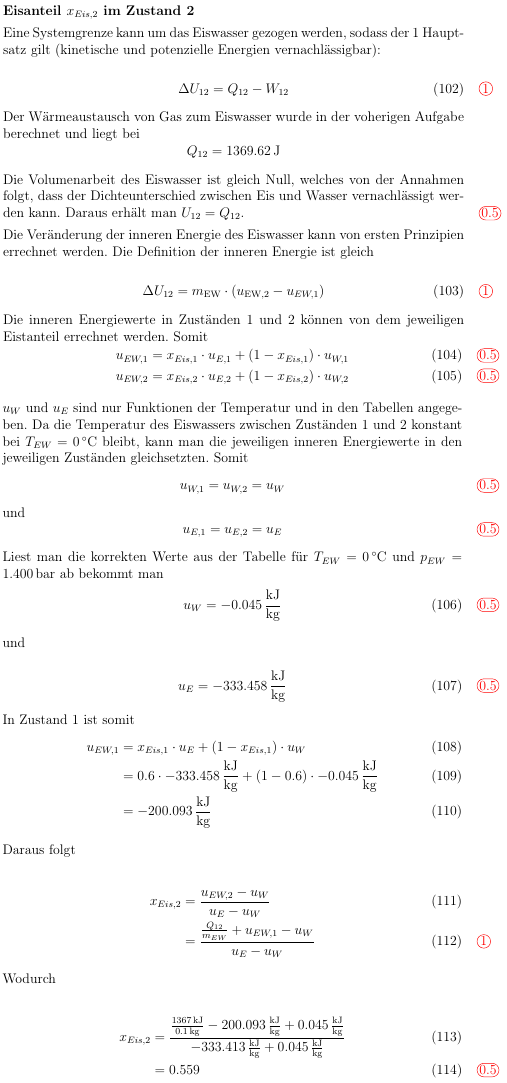}
\end{center}
\caption{Sample solution and rubric (in red) for Problem~3d (in the original German). Student solutions like the one in Fig.~\ref{fig:examplehand} would be graded based on this rubric.}
\label{fig:sample}
\end{figure}

An example of a sample solution is shown in Fig.~\ref{fig:sample}; generally, the student solutions were much shorter and far less explicit than the sample solution, see Fig.~\ref{fig:examplehand} (the synthetic data set used in an earlier study of automated grading by the authors~\cite{kortemeyer24aigrading} resembled the style of the sample solution rather than typical student solutions).

Problems~2 and~4 contain drawings. To grade those, GPT-4V was used to provide a textual description and embed it into the sample solution; Fig.~\ref{fig:graphdesc} demonstrates this for a process diagram. While the interpretation looks reasonable on the surface, and while certainly more accurate than the interpretation of the hand drawn diagram of the student solution in Fig.~\ref{fig:studentgraphdesc}, closer examination reveals a number of physics errors that would require manual intervention before being used in production settings.
 
\subsubsection{Problem-based Grading}
The system was prompted with the sample solution for that problem, and only the total number of points for the problem was queried. For the first problem, the system would simply return a value between 0~and 15~points.

\subsubsection{Grading Cycles}
Each of the MathPix-processed exams was graded ten times to compute the mean and standard deviation of the scores, resulting in 10 grades per exam component. For the subset of exams where five interpretations were produced by GPT-4V, each version was independently graded five times by GPT-4, resulting in a total of 25 grades per exam component. If a grading attempt failed because the LLM did not return one of the point values (depending on the granularity, one point value per rubric, one point value per part, or one point value per problem) or if any of these points values were out of range (e.g., assigning more than the available points), the grading cycle was repeated. Table~\ref{tab:expworkflows} gives an overview of the workflows that were explored (here named WF1 to WF4), including initial observations, which are discussed in more detail in the following sections.

\begin{table*}
\caption{Overview of explored grading workflows.\label{tab:expworkflows}}
\begin{ruledtabular}
\begin{tabular}{llllp{2cm}lp{2cm}llp{4cm}}

&Prompt	&Student	&Problem	&OCR	&\#OCR	&Grading	&Grading		&\#Grading 	&Initial Observations\\
&		&sample	&sample	&process	&cycles	&key		&granularity	&cycles		&\\
\hline
WF1&Fig.~\ref{fig:prompt_wf1}		&252	&1		&MathPix +~GPT-4V		&1	&Grading rubric items	&Rubric	&10	&Large number of failed grading attempts, necessitating frequent regrades.\\
WF2&Fig.~\ref{fig:prompt_wf2_wf4}	&252	&1--4	&MathPix +~GPT-4V		&1	&Whole sample solution	&Part	&10	&Less failed grading attempts.\\
WF3&Fig.~\ref{fig:prompt_wf3}		&252	&1		&MathPix +~GPT-4V		&1	&Whole sample solution	&Problem	&10	&Very few failed grading attempts. LLM still graded by part, but frequently ended up incorrectly adding up the points.\\
WF4&Fig.~\ref{fig:prompt_wf2_wf4}	&20	&1--4	&GPT-4V only			&5	&Whole sample solutions	&Part	&5	&Strong variations between OCR interpretations.\\
\end{tabular}
\end{ruledtabular}
\end{table*}

Figures~\ref{fig:prompt_wf1}--\ref{fig:prompt_wf3} show the AI prompts used for these workflows. These prompts may appear overly verbose to human readers and putting too much emphasis on obscure corner cases; unfortunately, they are the result of many initially failed attempts to provide proper grading output. It was also cumbersome to strike a balance between strict grading on the one side and being tolerant of what may or may not be OCR scanning errors.

During the grading steps, the grading script replaces the placeholders in square brackets by the respective documents (problem text, rubric or sample solution, and student solution) before submitting the prompt to the LLM. The output of the LLM is analyzed, and if it does not fulfill the requirements (missing rows, higher points than available, etc.), the same prompt is sent again. If the output is complete and passes all sanity tests, it is stored for calculating the total number of points and statistics.

\begin{figure*}[!ht]
\begin{flushleft}
\scriptsize
You are tasked with grading student solutions for this problem:

[problem]

==============

Below is the grading rubric table in comma-separated format. The first column contains the problem part, the second column provides
the rubric item identifier, the third column prescribes the maximum points possible for that item, and the fourth column contains
the grading criterion.

[rubric]

==============

The problem has multiple parts, and inside each part, it has several rubric items that are graded according to criteria.
Note that likely students will not have worked on all parts and items, and they may not have done work in the same order
as the rubric lists them.

This is what you need to do:

* Go through every row in the grading rubric table, do not skip any. The row is identified by the rubric item identifier listed in the
second column of the rubric table, and that is how you remember it. 

* For each row, check if any work pertaining to the criterion is present. If there is no related work, give zero points for that rubric item.

* For each row, if you find work that pertains to the criterion, check how well it corresponds to the criterion. Note that there can be
OCR errors, leading to some wrong digits or symbols; this is not the student's fault, and you need to determine if errors are likely due to OCR
(no point deduction) or due to the student (some point deduction). Give the maximum number of points listed for that row (third column
of rubric table) if the work corresponds very well. Give less points the less it agrees; you can use floating point numbers. Do not give bonus points. 

* For each row, provide a comment explaining the rationale behind your grading decision.

Output your grading in a four-column comma-separated table format, similar to the grading rubric table, with a row for every item in the rubric.
Use newline as the row separator. The first column in your table must be the problem part, the second column the rubric item identifier, the
third column the points that you awarded, and the fourth column your comments explaining the grading decisions. Output nothing else but this
completed grading table, and do not enclose it in any special characters, so it is syntactically correct CSV.

Here's the student solution that you need to grade:

[solution]
\end{flushleft}
\caption{Prompt for Workflow~1 (WF1 in Table~\ref{tab:expworkflows}). Square brackets indicate where the grading script will insert the respective documents before sending the prompt to the LLM.}
\label{fig:prompt_wf1}
\end{figure*}

\begin{figure*}[!ht]
\scriptsize
\begin{flushleft}
You are tasked with grading student solutions for this problem:

[problem]

==============

Below is the sample solution, provided by the instructors:

[sample]

==============

Here's the student solution that you need to grade:

[solution]

==============

You need to carefully compare each part of the student solution step-by-step to the respective part of the sample solution. The student answer
might not be in same order as the sample answer. Check carefully for completeness, as well as strictly for physics, math, and numerical accuracy.
Assign a percentage grade between 0 and 100 percent based on agreement with the sample solution, and prepare an explanation of your reasoning.
Reserve high percentages for excellent work, i.e., almost perfect agreement with the sample solution. Do not hesitate to give very low percentages
for incomplete, incoherent, or inaccurate work.

Output the grade percentage for each part as an integer between 0 and 100 and then in quotation marks provide an explanation of your reasoning
for this grade in the form

part,percentage,explanation
part,percentage,explanation
...

Output nothing else but this list.
\end{flushleft}
\caption{Prompt for Workflows~2 and~4 (WF2 and WF4 in Table~\ref{tab:expworkflows}).}
\label{fig:prompt_wf2_wf4}
\end{figure*}

\begin{figure*}[!ht]
\scriptsize
\begin{flushleft}
You are tasked with grading student solutions for this problem:

[problem]

==============

Below is the sample solution, provided by the instructors:

[sample]

==============

Here's the student solution that you need to grade:

[solution]

==============

You need to carefully compare the student solution step-by-step to the sample solution.

The student answer might not be in same order as the sample answer. Check carefully for completeness, as well as strictly for physics, math, and numerical accuracy.
Assign a points based on agreement with the sample solution, and prepare an explanation of your reasoning. Reserve full points for excellent work, i.e., almost perfect
agreement with the sample solution. Do not hesitate to give very few points for incomplete, incoherent, or inaccurate work.

Output the points and then in quotation marks provide an explanation of your reasoning for this grade in the form, all in one single line prefaced by "sum", i.e.,

"sum",points,explanation

Output nothing else but this one line.
\end{flushleft}
\caption{Prompt for Workflow~3 (WF3 in Table~\ref{tab:expworkflows}).}
\label{fig:prompt_wf3} 
\end{figure*}

\subsection{Evaluation}
For the evaluation, the exam points awarded by the TAs were assumed as ground truth. Evaluations were carried out using Python scripts, Excel, and R~\cite{rproject}. Taking the standard deviations as a measure of confidence, agreement with TA grading was evaluated for different confidence levels using linear regression and Spearman correlations.

\section{Results}

\subsection{Rubric-based grading based on MathPix and GPT-4V (WF1)}
\begin{table*}
\caption{Example of grading the student solution in Fig.~\ref{fig:prob1hand} according to the rubric in Table~\ref{tab:grading_rubric}, following WF1}
\begin{tabular}{llllllllllllll}
Item&AI~1&AI~2&AI~3&AI~4&AI~5&AI~6&AI~7&AI~8&AI~9&AI~10&AI~Ave.&AI-SD&TA\\
\hline
a\_Bilanz&0.5&0.8&0.6&1&1&1&1&1&1&0.8&0.87&0.179&1\\
a\_h\_ein&0.5&0.5&0.5&0.5&0.5&0.5&0.5&0.5&0.5&0.5&0.5&0.000&0.5\\
a\_h\_aus&0&0&0&0.5&0.5&0.5&0&0.5&0&0.5&0.25&0.250&0.5\\
a\_Ergebnis&0&0.5&0.3&0.5&0.3&0.5&0.5&0.3&0.5&0.5&0.39&0.158&0\\
b\_T\_KF\_Tds&0.5&0.5&0.8&1&1&1&1&0.8&1&1&0.86&0.196&1\\
b\_Stoffmod&0.5&0.5&1&1&1&1&1&0.5&1&1&0.85&0.229&1\\
b\_Erg&0&0.5&0.5&0&0.3&0.5&0.5&0.35&0.5&0.4&0.355&0.190&0.5\\
c\_Bilanz&0.5&0.5&1&1&1&1&1&1&1&1&0.9&0.200&1\\
c\_Erg&0&0.5&0.5&0.5&0.5&0.5&0.5&0.5&0.5&0&0.4&0.200&0.5\\
d\_u\_f&0&0.5&0&0.5&0.5&0.5&0.5&0.45&0&0&0.295&0.241&0.5\\
d\_u\_g&0.5&0.5&0&0.5&0.5&0.5&0.5&0.45&0&0&0.345&0.226&0.5\\
d\_u\_1&0.5&0.5&0.5&1&1&1&1&0.7&0&1&0.72&0.325&1\\
d\_u\_2&0&0&0&0.5&0.5&0.5&0.5&0.3&0.5&0.5&0.33&0.224&0.5\\
d\_h\_ein&0&0.5&0.5&0.5&0.5&0.5&0.5&0.5&0.5&0.5&0.45&0.150&0.5\\
d\_Bilanz&1&1&1&1&0.7&1&1&1&0.5&1&0.92&0.166&0\\
d\_Erg&0&0.5&0.5&0.5&0.3&0.5&0.5&0.4&0.5&0.5&0.42&0.154&0\\
e\_s\_f&0&0.5&0&0.5&0.5&0.5&0.5&0.5&0&0.5&0.35&0.229&0.5\\
e\_s\_g&0&0.5&0&0.5&0.5&0.5&0.5&0.5&0&0.5&0.35&0.229&0.5\\
e\_s\_1&0.5&0.5&0.5&1&1&0.5&0&0.75&1&1&0.675&0.317&1\\
e\_s\_2&0&0.5&0&0.5&0.5&0.5&0&0.5&0.5&0.5&0.35&0.229&0.5\\
e\_Bilanz&0.5&0.5&0&1&1&1&0&0.6&1&1&0.66&0.388&0\\
e\_Erg&0&0&0.2&0.5&0.4&0.5&0&0.35&0.5&0.5&0.295&0.213&0\\
&&&&&&&&&&&&&\\
Total&5.5&10.3&8.4&14.5&14&14.5&11.5&12.45&11&13.2&11.535&2.745&11.5

\end{tabular}

\label{tab:grad_rubr}
\end{table*}
Table~\ref{tab:grad_rubr} shows the outcome of grading the student solution in Fig.~\ref{fig:prob1hand} according to the rubric in Table~\ref{tab:grading_rubric}. The rows are the grading items with the final row showing the total points. The columns are the results of ten AI-runs, followed by the average of the AI grades, the standard deviation of the AI grades, and the TA grades as ground truth. Here, the Law of Large numbers seems to have applied, as the grading outcomes of the AI and the TAs are almost identical; this, unfortunately, is rarely the case.

\begin{figure*}
\begin{center}
\includegraphics[width=0.76\textwidth]{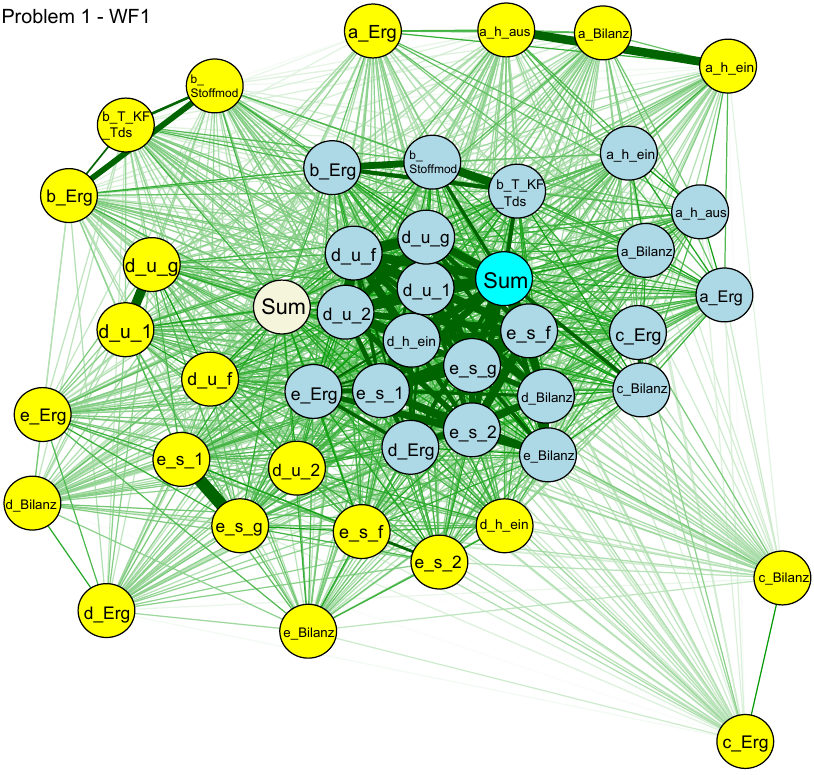}
\end{center}
\caption{Fruchterman-Reingold~\citep{fruchterman1991,qgraph} representation of all Spearman correlations between the rubric items in Table~\ref{tab:grading_rubric}. Indicated in yellow are the grades given by the TAs, indicated in blue those by the AI. The thickness of the edges indicates the strength of the Spearman correlation, where negative correlations would have been indicated in red. The beige ``Sum'' and cyan ``Sum'' indicate the sum of points given by the teaching assistants and the AI, respectively.}
\label{fig:rubriccor}
\end{figure*}

Figure~\ref{fig:rubriccor} shows a Fruchterman-Reingold~\citep{fruchterman1991,qgraph} representation of the Spearman-correlation matrix between the rubric items in Table~\ref{tab:grading_rubric} for Problem~1, resulting from Workflow~1 (WF1) in Table~\ref{tab:expworkflows}. Indicated in yellow are the rubric grades given by the TAs, indicated in blue those given by the AI. Due to the force-directed nature of Fruchterman-Reingold graphs, closely correlated vertices tend to cluster, while unrelated vertices tend to be further apart from each other.

An immediate observation in the graph is that the AI-grades cluster together much more strongly than those of the TAs, which could be explained by the TAs being able to grade the items independently, while the AI tends to grade holistically with only limited ability to differentiate between one step of a solution being correct and another one incorrect. For example, for the TAs, e\_Bilanz and e\_s\_2 are weakly correlated (lower middle of the graph), while in the AI grading, they are strongly correlated.

For both TA and AI grades, however, a clustering by parts a--e  is observable, indicating that problem parts tend to be correct or incorrect as a whole (again, this effect is stronger for the AI than the TA grades). While generally within each other's vicinity, pairs of TA and AI grades for the same rubric items are not strongly correlated.

The beige vertex ``Sum'' indicates the total grade for the problem given by the TAs (``sum of the yellow vertices''), while the corresponding cyan vertex indicates the same for the AI grades. Remarkably, this pair of summative vertices is more  closely aligned than many of the individual pairs of rubric items. The holistic, undifferentiated tendencies of the AI might be mediated by students tending to get the whole problem mostly correct or mostly incorrect as a whole, based on their understanding of the underlying subject matter.

\begin{figure*}
\begin{center}
\includegraphics[width=0.48\textwidth]{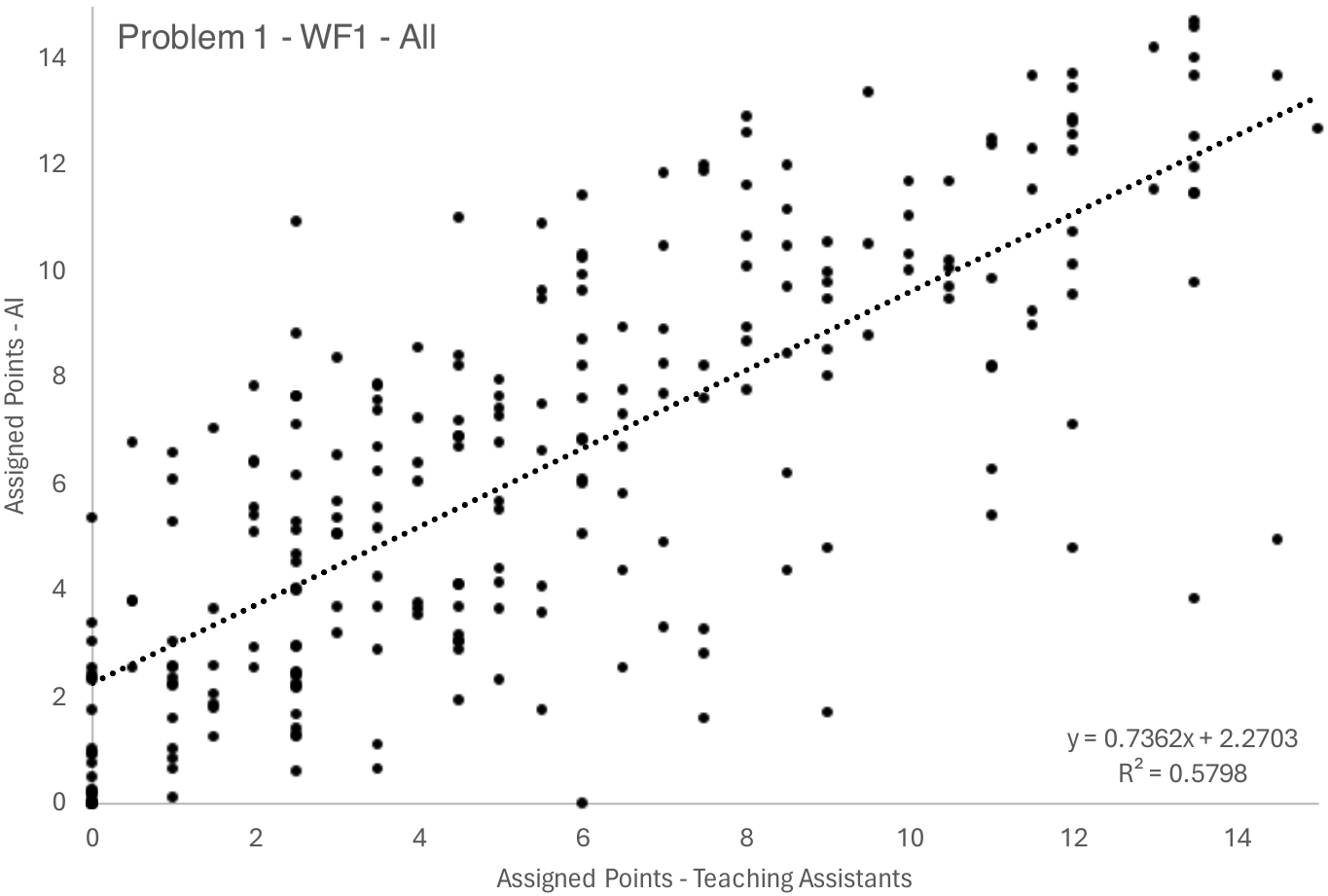}
\includegraphics[width=0.48\textwidth]{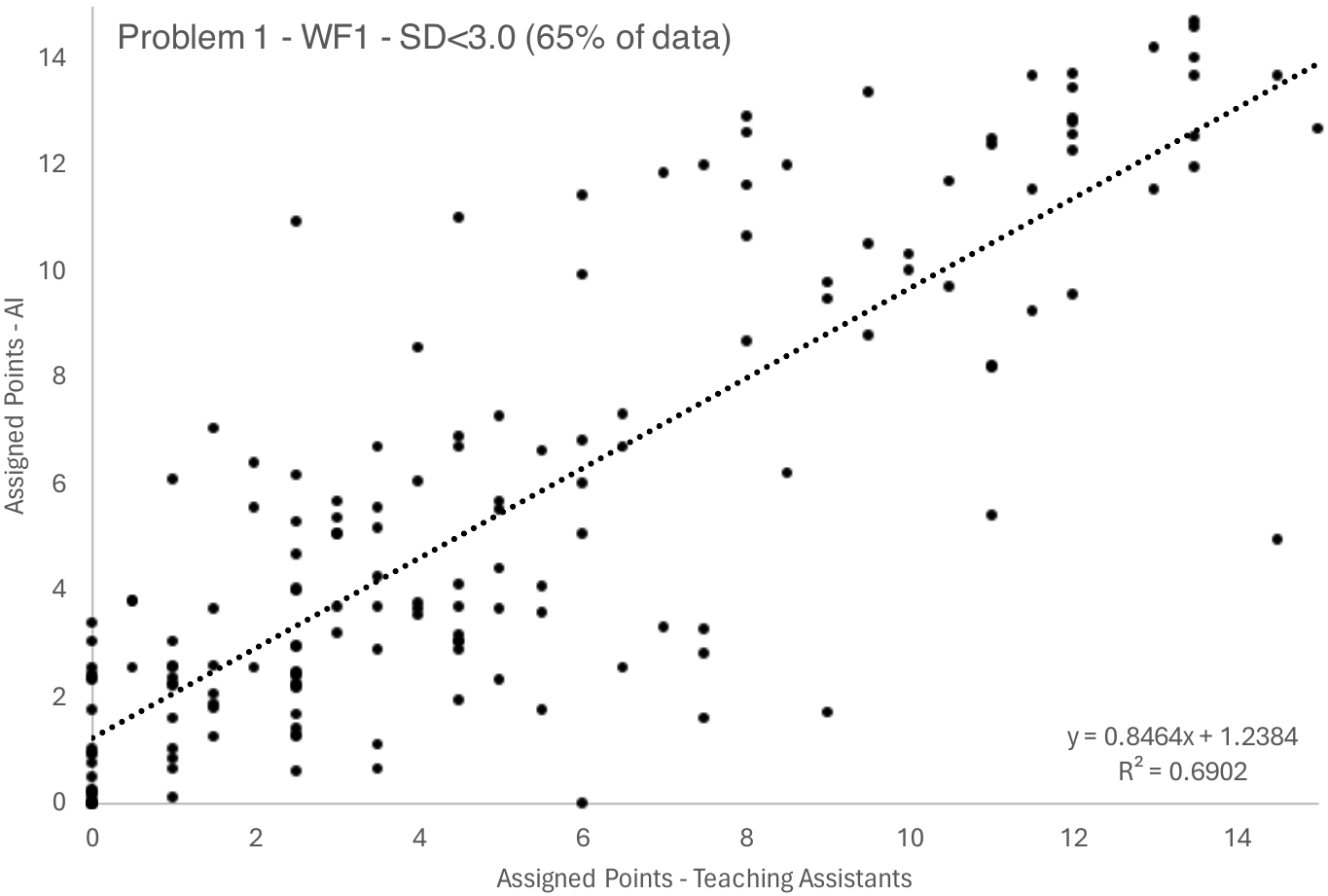}

\includegraphics[width=0.48\textwidth]{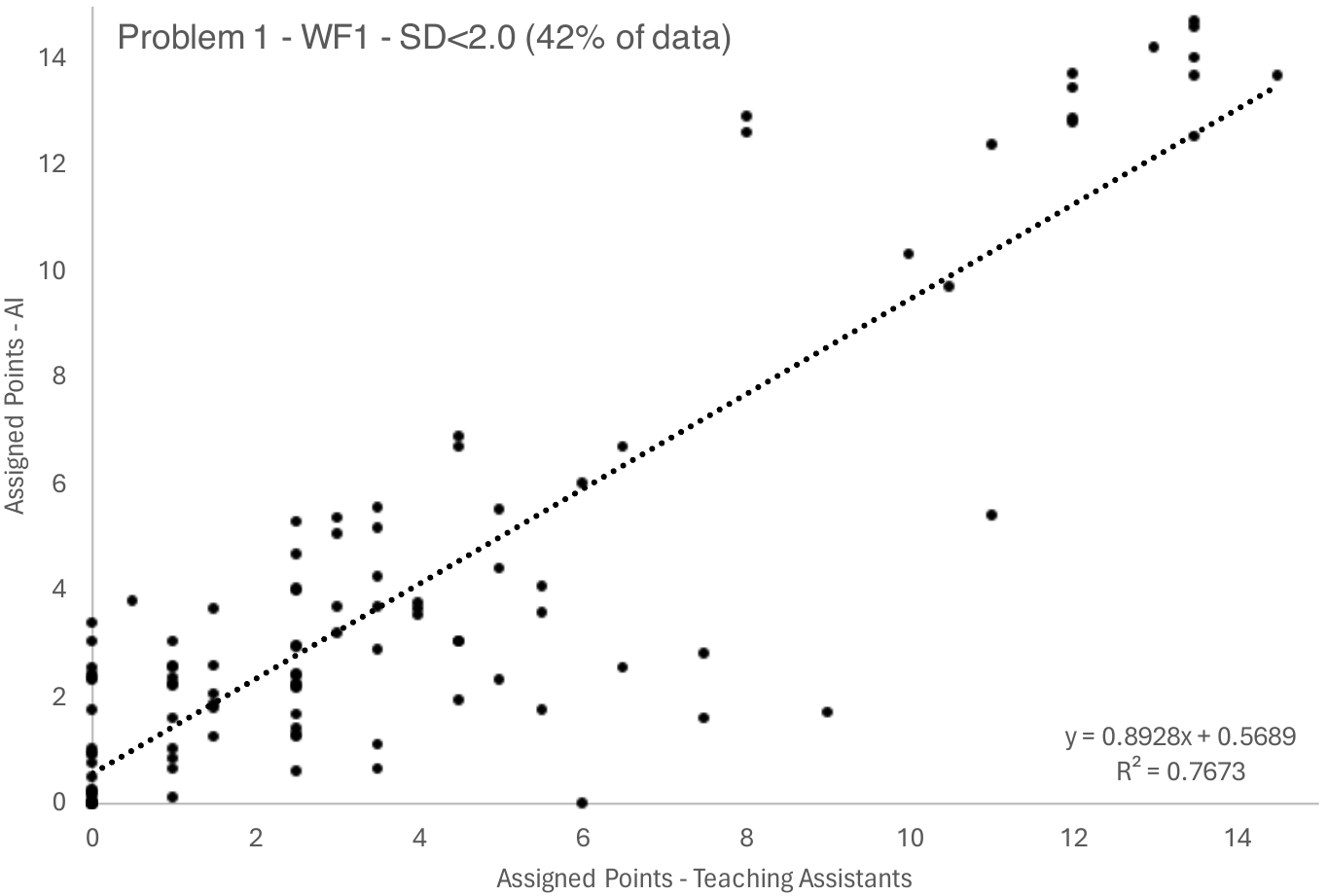}
\includegraphics[width=0.48\textwidth]{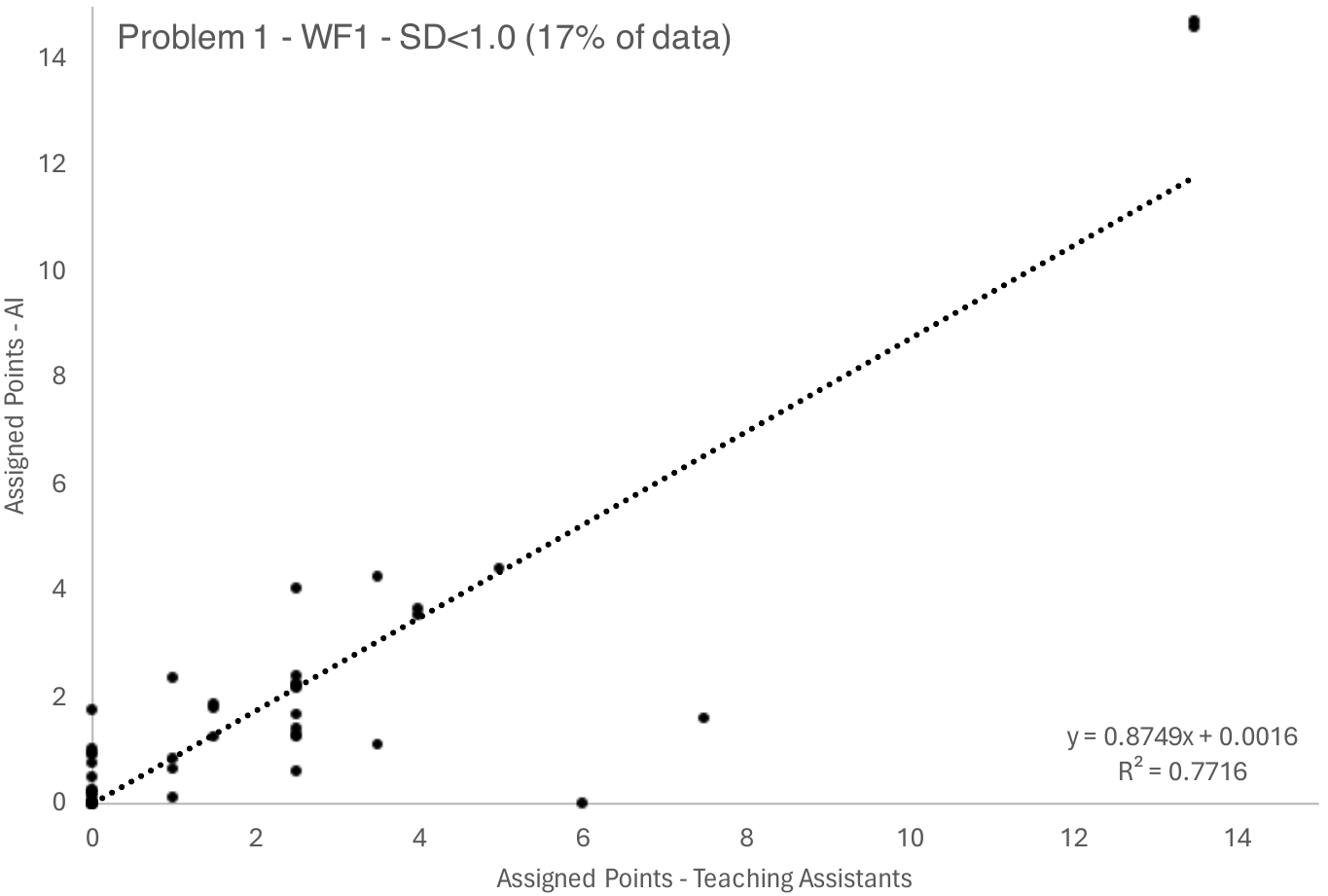}
\end{center}
\caption{AI versus TA grades for Problem~1 using WF1 including linear regression lines, their equations and $R^2$. The upper left panels shows all grades independent of standard deviation $\sigma$ of the mean of the 10~AI grades, the remaining panels the relationship for increasingly stringent limitations on the standard deviation. In each case, the percentage of data points fulfilling the restriction is indicated.}
\label{fig:rubricgrades}
\end{figure*}

Figure~\ref{fig:rubricgrades} illustrates the relationship between the total grades for Problem~1 given by the TAs and the AI. Each data point represents one exam. Considering all grading results, the coefficient of determination is only $R^2=0.58$, which would generally be considered moderate. When demanding standard deviations on the AI grading of maximal three, two, or one point, the percentage of AI-gradings fulfilling these conditions decreases while the coefficient of determination $R^2$ moves into the direction of what might be considered strong. Problems below the set confidence level would need to be graded by hand.

Notably, outliers with very low AI-points remain even at high confidence thresholds; these are due to OCR issues, where MathPix and GPT-4V were unable to interpret the handwriting, and thus  the AI-graders consistently gave low points due to missing information. This entails that regardless of the set confidence threshold,  problems with very low points should be graded or at least checked by humans.

As observed in earlier studies, in spite of the prompts in Figs.~\ref{fig:prompt_wf1}--\ref{fig:prompt_wf3} attempting to enforce strict grading, the AI grades more leniently than TAs~\cite{kortemeyer24aigrading}, but the $R^2=0.84$ achieved with the more verbose synthetic answers could not be reproduced with authentic data.

The low agreement between TAs and AI on individual rubric items, as well as the very frequent failures to provide a complete grading output with points within the available range, indicates that fine-grained rubric grading for a complete problem is beyond  GPT-4-32k's ``bookkeeping'' capabilities.The high computational load caused by the frequent need to regrade makes WF1 appear nonviable. A variant of this workflow might be prompting the LLM for one part at a time, only submitting the handful of rubric items pertaining to that part; the LLM would likely be able to keep track of this limited number of grading items, but this approach, for this exam, would have come at the expense of four times more LLM transactions.

\subsection{Part-based grading based on MathPix and GPT-4V (WF2)}
\begin{figure*}
\begin{center}
\includegraphics[width=0.48\textwidth]{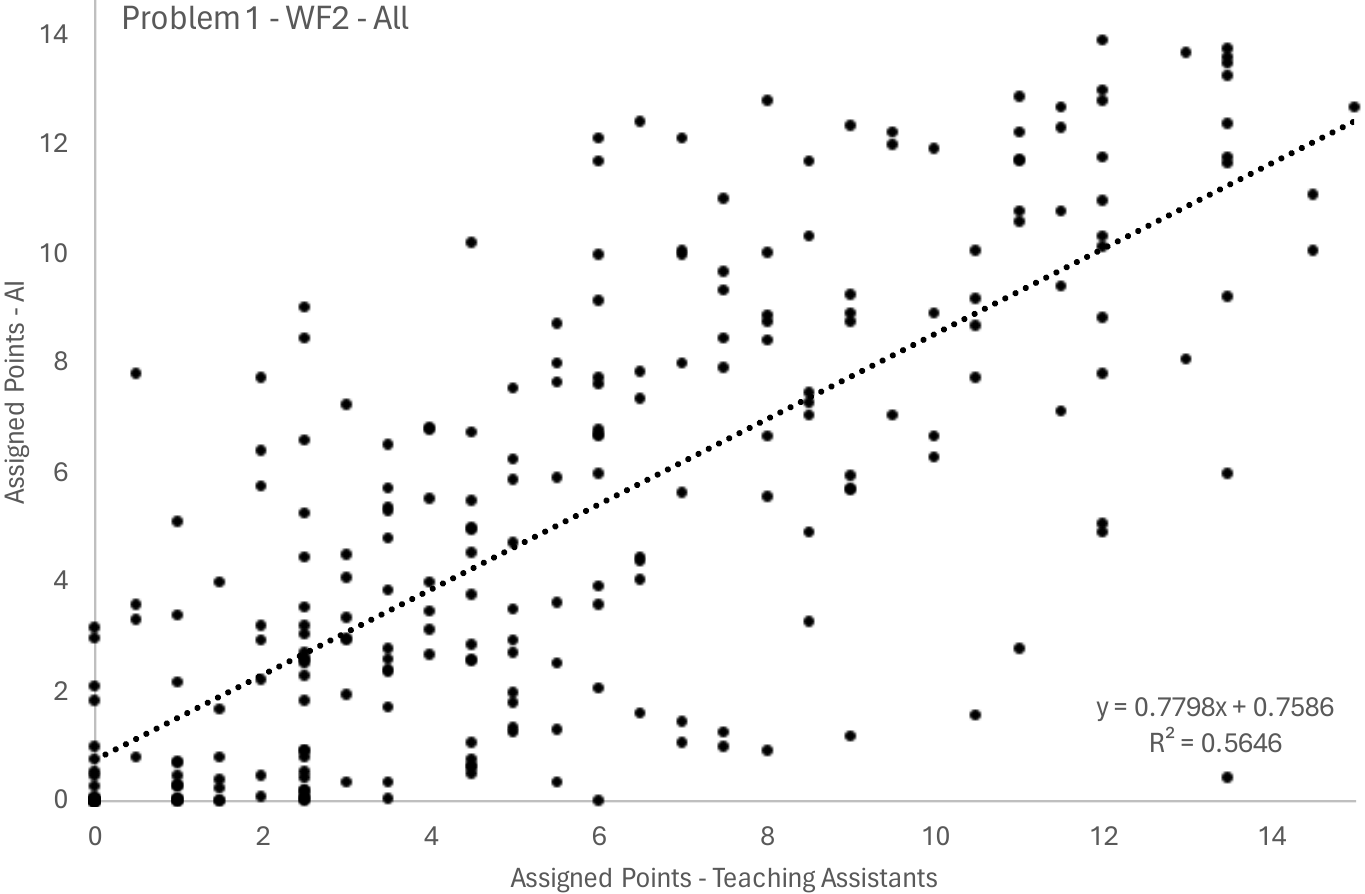}
\includegraphics[width=0.48\textwidth]{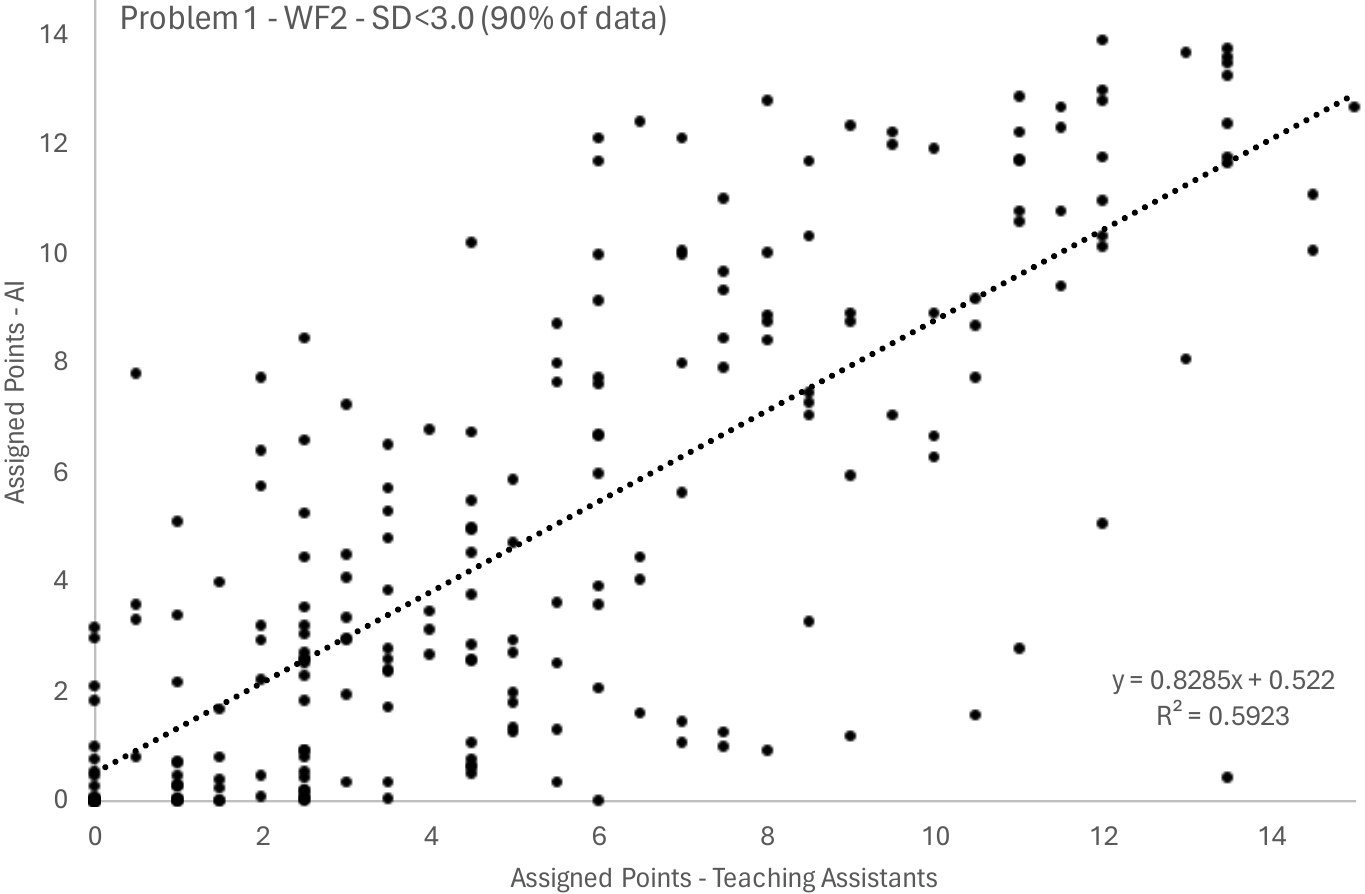}

\includegraphics[width=0.48\textwidth]{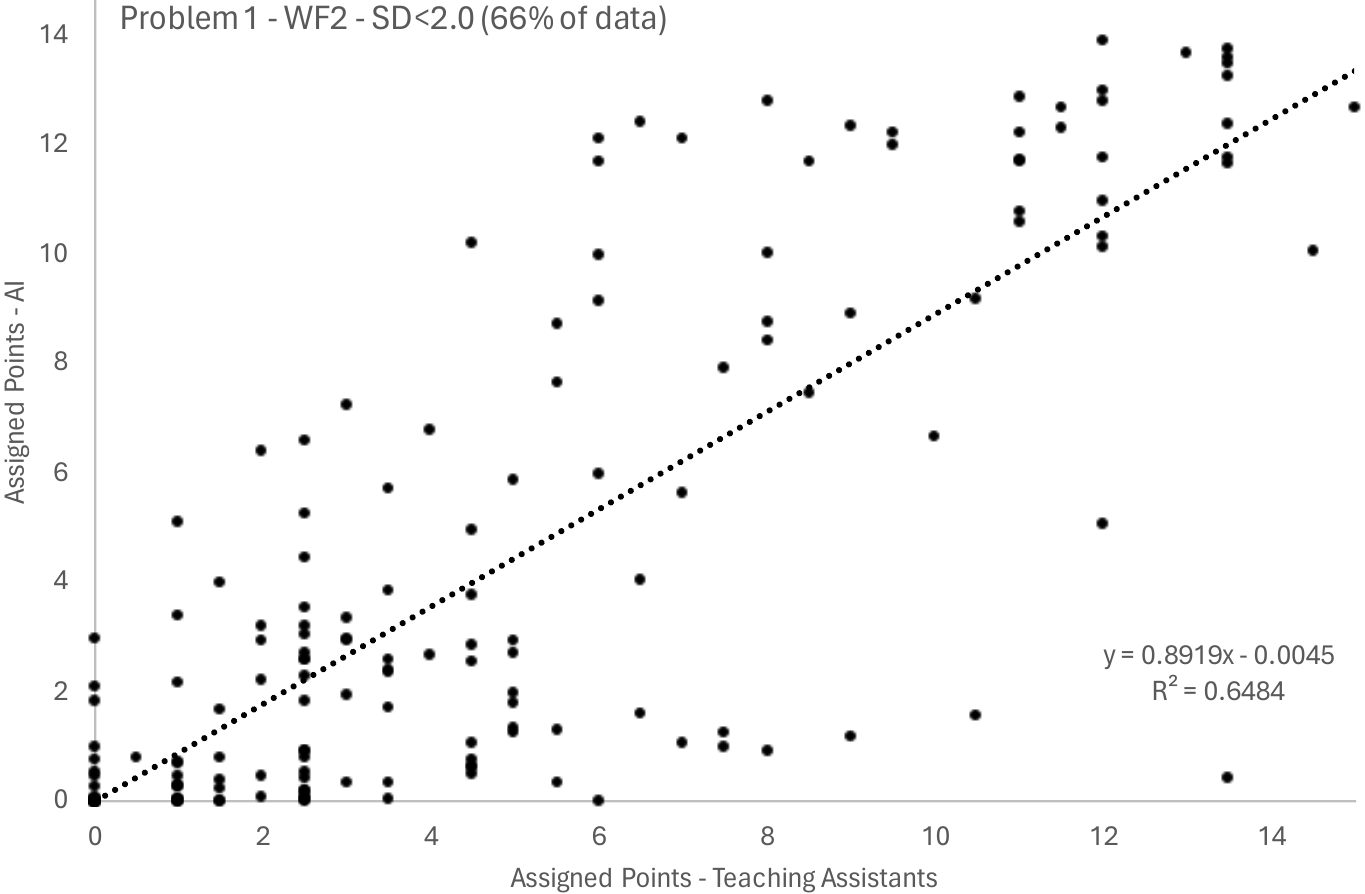}
\includegraphics[width=0.48\textwidth]{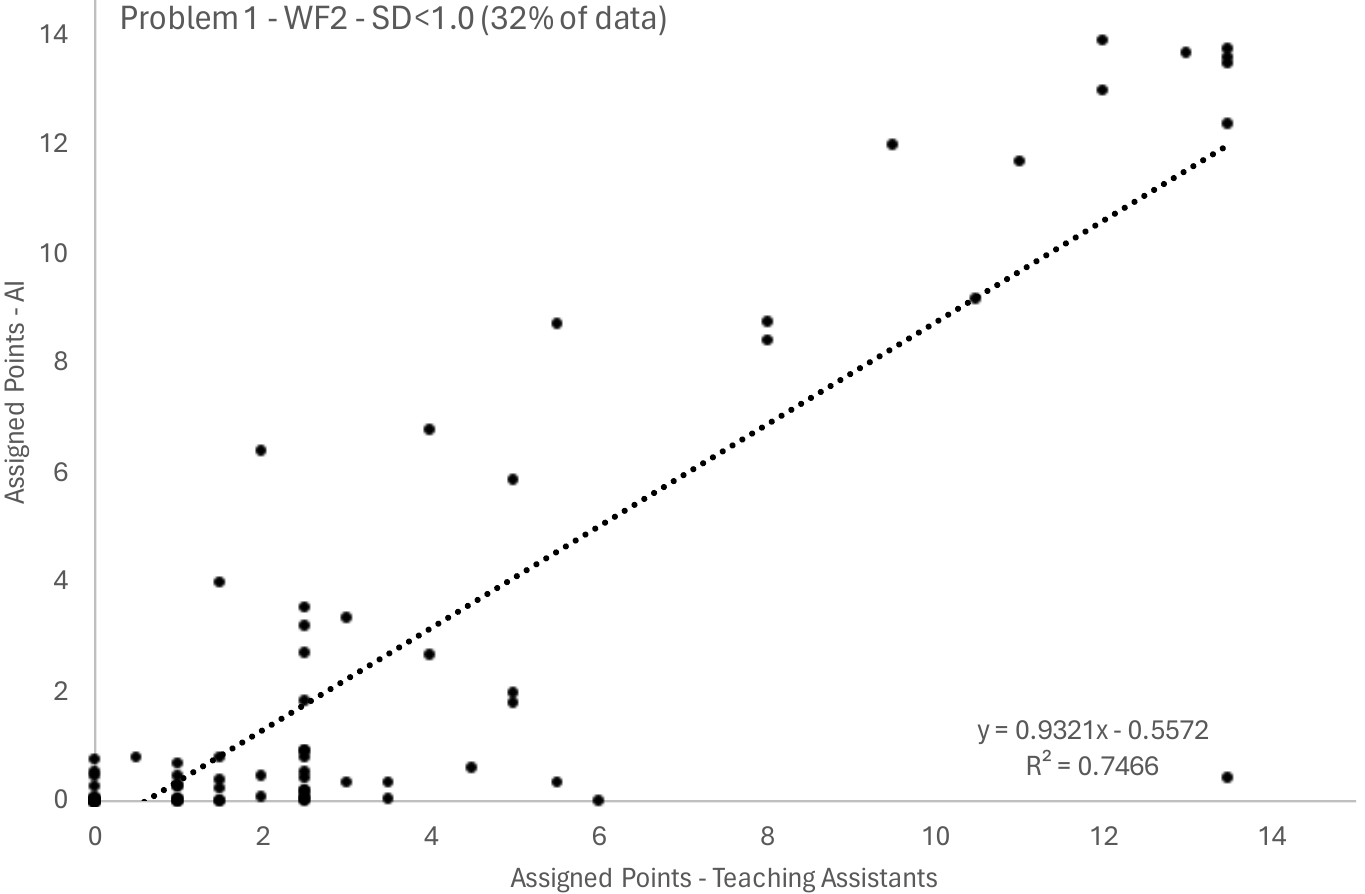}
\end{center}
\caption{AI versus TA grades for Problem~1 using WF2 including linear regression lines, their equations and $R^2$.}
\label{fig:wholegrades}
\end{figure*}

\begin{table*}
\caption{Example of grading the student solution in Fig.~\ref{fig:prob1hand} based on the sample solution by parts, following WF2}
\begin{tabular}{llllllllllllll}
Part&AI~1&AI~2&AI~3&AI~4&AI~5&AI~6&AI~7&AI~8&AI~9&AI~10&AI~Ave.&AI-SD&TA\\
\hline
a&2&2.25&2&2&2.375&2.125&1.5&2&2&2.25&2.05&0.225&2\\
b&2.125&2.5&1.75&2.25&2.125&2.125&2.5&2.25&2&2.5&2.2125&0.231&2.5\\
c&1.35&1.5&1.2&1.5&1.35&1.35&1.35&1.35&1.5&1.425&1.3875&0.090&1.5\\
d&3.375&4.5&3.15&3.825&3.6&3.825&3.15&3.15&3.6&4.05&3.6225&0.420&3\\
e&2.8&3.6&2.8&2.8&3.2&3.4&2.2&2.8&2.8&4&3.04&0.488&2.5\\
&&&&&&&&&&&&&\\
Total&11.65&14.35&10.9&12.375&12.65&12.825&10.7&11.55&11.9&14.225&12.3125&1.182&11.5

\end{tabular}

\label{tab:grad_part}
\end{table*}
Workflow~2 (WF2) considers problem parts rather than rubric items. Initial tests on Problem~1 showed that the LLM was much better equipped to keep track of four or five problem parts than of 22~rubric items, and the number of necessary regrades dramatically decreased; Table~\ref{tab:grad_part} shows an example for the problem solution in Fig.~\ref{fig:prob1hand}. The coefficient of determination for TA versus AI grades for Problem~1 was $R^2=0.56$ for the complete data set and thus comparable to the $R^2=0.58$ obtained from rubric grading, see Fig.~\ref{fig:wholegrades}. The standard deviations of the average grades of the 10~AI graders are lower than in workflow WF1, which results in less exams requiring human grading when imposing limits on the maximum standard deviation.

\begin{figure*}
\begin{center}
\includegraphics[width=0.36\textwidth]{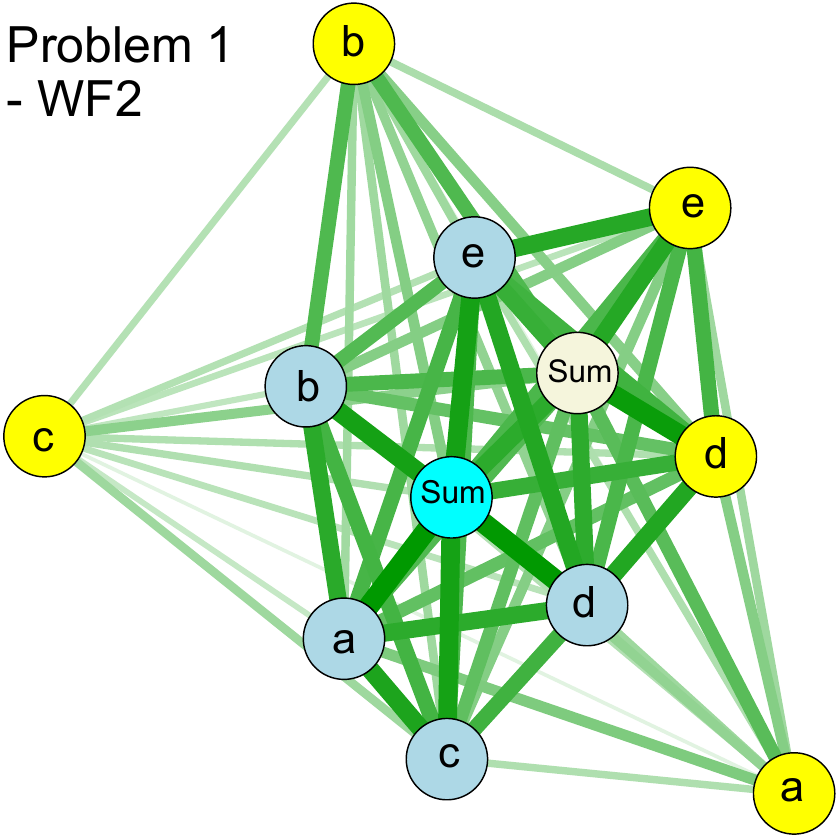}\qquad
\includegraphics[width=0.36\textwidth]{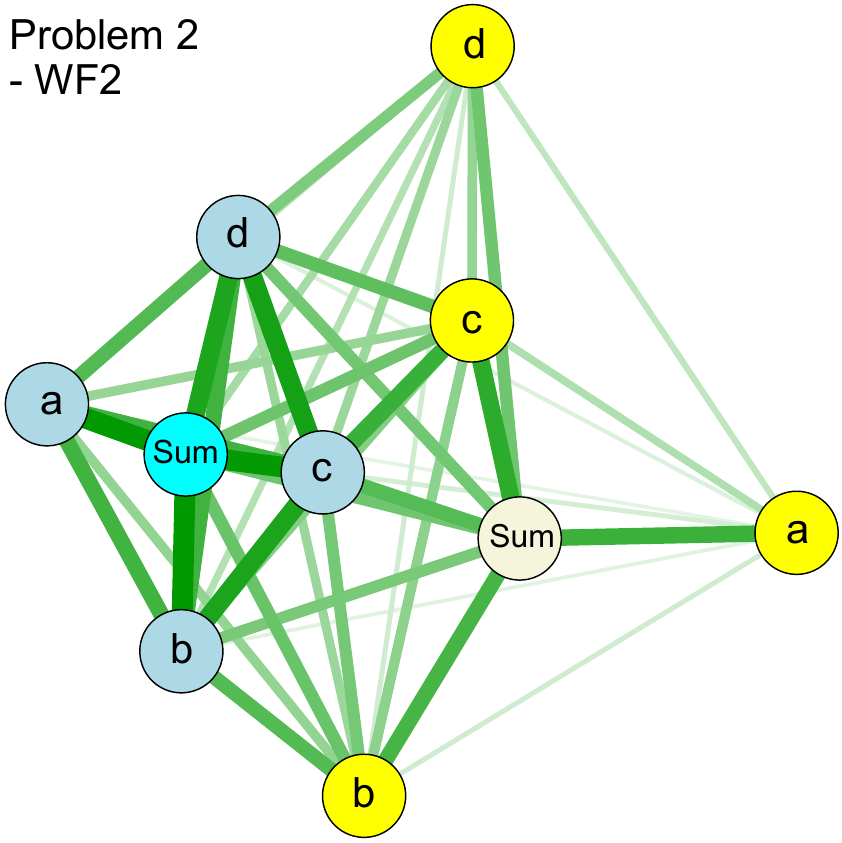}

\includegraphics[width=0.36\textwidth]{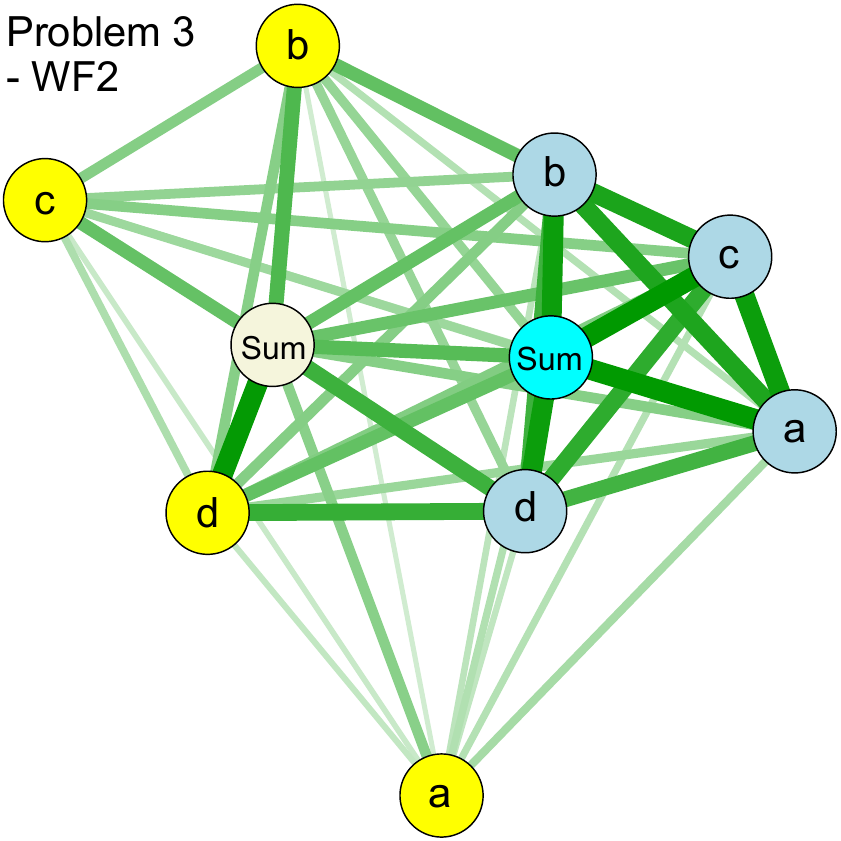}\qquad
\includegraphics[width=0.36\textwidth]{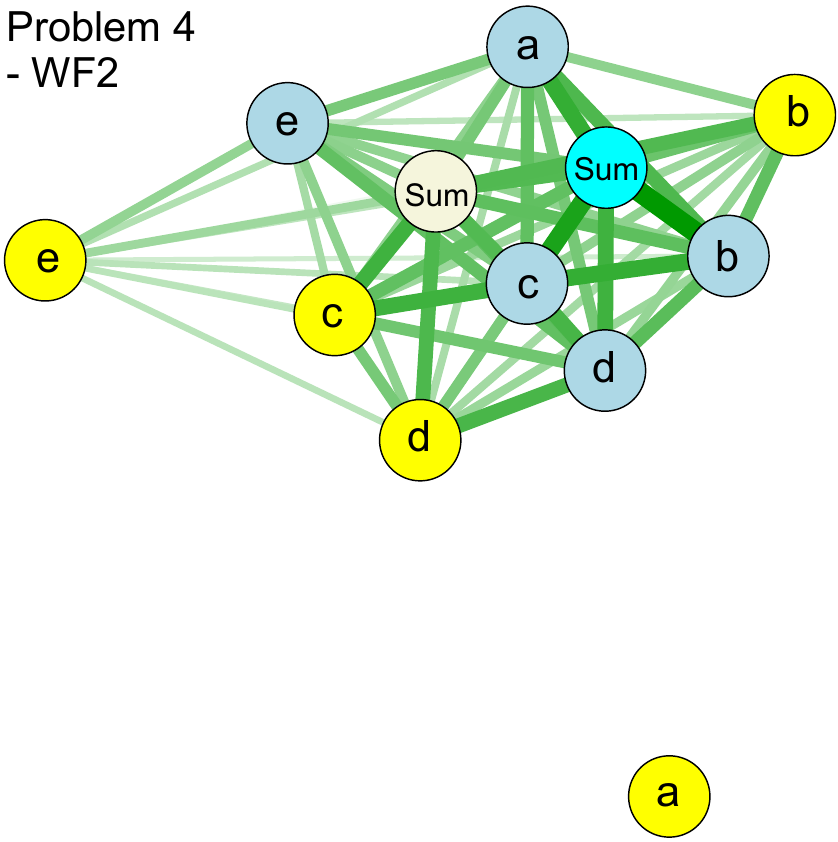}
\end{center}
\caption{Fruchterman-Reingold~\citep{fruchterman1991,qgraph} representation of all Spearman correlations between the part grades for all problems on the exam. Indicated in yellow are the grades given by the TAs, indicated in blue those by the AI. The beige ``Sum'' and cyan ``Sum'' indicate the sum of points given by the teaching assistants and the AI, respectively.}
\label{fig:allpartgrades}
\end{figure*} 

For this workflow, all problems were considered, and Fig.~\ref{fig:allpartgrades} shows the Spearman correlations between the TA and AI grades.  As seen earlier in Fig.~\ref{fig:rubriccor}, the sums of the parts are more closely correlated than the pairs of parts. Very prominent is the outlying vertex for the TA-graded Part~4a, which is one of two graphical problem parts. As opposed to the other graphical problem part, Part~2a, which contributes $8$~points to the $19$ points of Problem~2 (explaining the strong correlation with ``Sum''), Part~4a only contributes $4.5$~points to the $18.5$~points of Problem~4.  The AI-graded Parts~2a and~4a, on the other hand, are simply part of the cluster of AI-points visible in all Spearman correlations.

The pairs of vertices for the graphical problems~2a and~4a show no correlations (for 2a, the vertices are at opposite ends of the graph, while for 4a, the TA-vertex is completely disconnected). These would be grades resulting from comparing the descriptions of solutions like Fig.~\ref{fig:studentgraphdesc} to those in Fig.~\ref{fig:graphdesc}, which is a task that LLMs would in principle be capable of if the descriptions were reliable; unfortunately, the latter proved not to be the case.

\subsection{Problem-based grading based on MathPix and GPT-4V (WF3)}
\begin{table*}
\caption{Example of grading the student solution in Fig.~\ref{fig:prob1hand} based on the sample solution, not prompting for parts, following WF3}
\begin{tabular}{llllllllllllll}
&AI~1&AI~2&AI~3&AI~4&AI~5&AI~6&AI~7&AI~8&AI~9&AI~10&AI~Ave.&AI-SD&TA\\
\hline
Total&11.5&10&12.5&12.5&10&11.5&10&11&9&13&11.1&1.261&11.5
\end{tabular}

\label{tab:grad_problem}
\end{table*}

Grading a complete problem with only one summary grade turned out to have the lowest correlation with TA grades of all workflows, $R^2=0.43$. Table~\ref{tab:grad_problem} shows an example for the student solution in Fig.~\ref{fig:prob1hand}. The grading explanations showed that the model still attempted to grade by part, for example it graded the solution in Fig.~\ref{fig:prob1hand} in Run AI1 (see Table~\ref{tab:grad_problem}) as follows:
\begin{quote}\footnotesize
The student solved parts a), b) and c) correctly, earning 2.5 points, 2.5 points and 1.5 points respectively. For part d), the student made a miscalculation resulting in a discrepancy with the model solution, earning 2 of the 4.5 points. For part e), the student made another miscalculation resulting in a discrepancy with the model solution, earning 2.5 of the 4 points.
\end{quote}
The model extracted the point values for the parts from the problem text, which was part of the prompt (see Fig.~\ref{fig:prompt_wf3}). However, it was unable to correctly add five numbers, as 2.5+2.5+1.5+2+2.5 is 11, but the model reported 11.5 in the output. As the model seems to proceed by parts anyway, but cannot be trusted to add up the points correctly, it is seems advisable to leave arithmetics to Python or Excel. 
\subsection{Part-based grading based solely on GPT-4V (WF4)}
\begin{table*}
\caption{Example of grading the student solution in Fig.~\ref{fig:prob1hand} based on the sample solution by parts, following WF4}
\begin{tabular}{llllllllllllllllll}
Part&AI~1.1&AI~1.2&AI~1.3&AI~1.4&AI~1.5&AI~2.1&\ldots&AI~4.5&AI~5.1&AI~5.2&AI~5.3&AI~5.4&AI~5.5&AI~Ave.&AI-SD&TA\\
\hline
a&0&1&0&1.875&0.25&1.5&&0.25&2.125&1.75&1.5&2&1.75&0.935&0.760&2\\
b&0&0.75&1.25&1.125&0&1.25&&0&2.25&1.5&2.25&2.5&2.25&0.9&0.846&2.5\\
c&1.05&1.35&1.125&1.275&0.225&1.35&&0&1.425&1.2&1.425&1.35&1.5&0.729&0.601&1.5\\
d&3.375&2.25&2.25&4.05&0&0&\ldots&1.8&2.7&2.7&2.25&3.825&2.7&1.989&1.452&3\\
e&3.4&4&3.2&2.2&0.6&0&&0&2&2.8&2.4&4&2&1.76&1.397&2.5\\
&&&&&&&&&&&&&&&&\\
Total&7.825&9.35&7.825&10.525&1.075&4.1&\ldots&2.05&10.5&9.95&9.825&13.675&10.2&6.313&3.912&11.5
\end{tabular}

\label{tab:gpt_grad_part}
\end{table*}

Workflow WF4 has multiple GPT-4V-only interpretations of the handwriting (see Fig.~\ref{fig:ocr}) and multiple GPT-4-32k graders, resulting in more than double of the grades generated by the other workflows (25~versus~10).  This choice was made in the hope of decreasing standard deviations on the average grades, however, the workflow turned out to have higher standard deviations than WF2 and WF3. Table~\ref{tab:gpt_grad_part} shows an example for the solution in Fig.~\ref{fig:prob1hand}. As increasingly strict limits on the standard deviation are imposed, the Coefficient of Determination $R^2$ did not consistently increase. However, due to the high computational load, this approach was abandoned after processing 20~sample exams, so this lack of consistency might be the result of both workflow-inherent uncertainties in the OCR process and simple lack of statistics (20~versus~252 data points).

Also in this workflow, student graphs like the one in Fig.~\ref{fig:studentgraphdesc} were frequently ``overlooked,'' and grading results of the recognized graphs were unreliable. GPT-4V seems much better equipped to describe graphs with clean lines like the one in Fig.~\ref{fig:graphdesc} than freehand drawings like Fig.~\ref{fig:studentgraphdesc}.

\subsection{Influence of Confidence Thresholds}

\begin{figure}
\begin{center}
\includegraphics[width=\columnwidth]{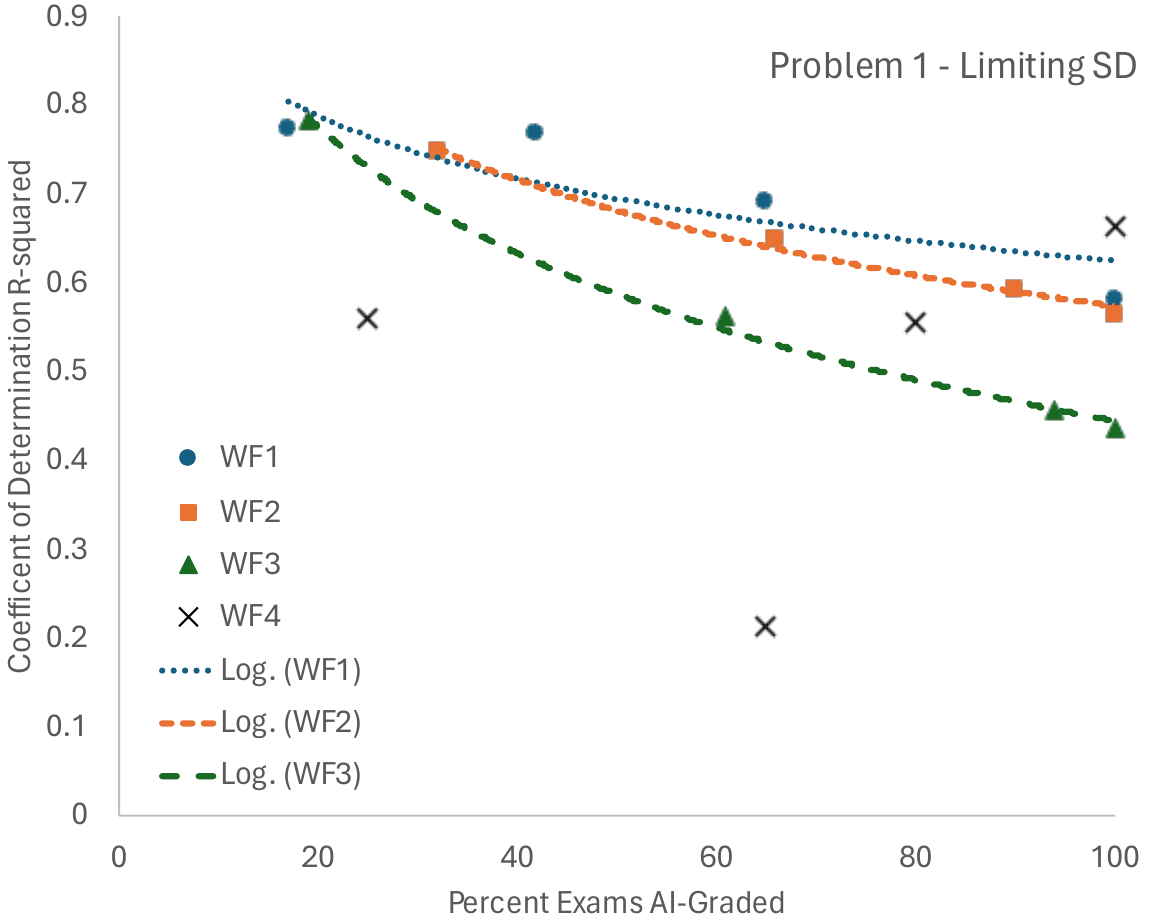}
\end{center}
\caption{Coefficient of determination $R^2$ versus percentage of accepted AI-grades for Problem~1 as restrictions are put on the standard deviation $\sigma$ of the AI-grades, comparing the different workflows.}
\label{fig:compareWFs}
\end{figure} 

Figure~\ref{fig:compareWFs} summarizes the results of Coefficient of Determination $R^2$ versus imposed limits on the standard deviation. For W1-WF3, logarithmic fits are included; for WF4, the randomness is too high or the statistics too low to provide a reasonable fit. From left to right, the data points represent acceptance of standard deviations lower than one, three, or any number of points, respectively; thus, data points further to the left on the horizontal axis indicate greater need for human grading, while the values on the vertical axis indicate better agreement between TA and AI grading.

WF4 does not seem ready for production, as it is more unpredictable than the other workflows.
Workflows WF1 and WF2 clearly outperform WF3 (possibly simply because GPT-4-32k cannot add numbers). If, based on confidence measures, half of the exams are AI-graded, both models reach $R^2\approx0.7$. While these coefficients indicate strong correlations, it is important to note that they would be unacceptable for high-stakes exams. 

\begin{figure*}
\begin{center}
\includegraphics[width=0.49\textwidth]{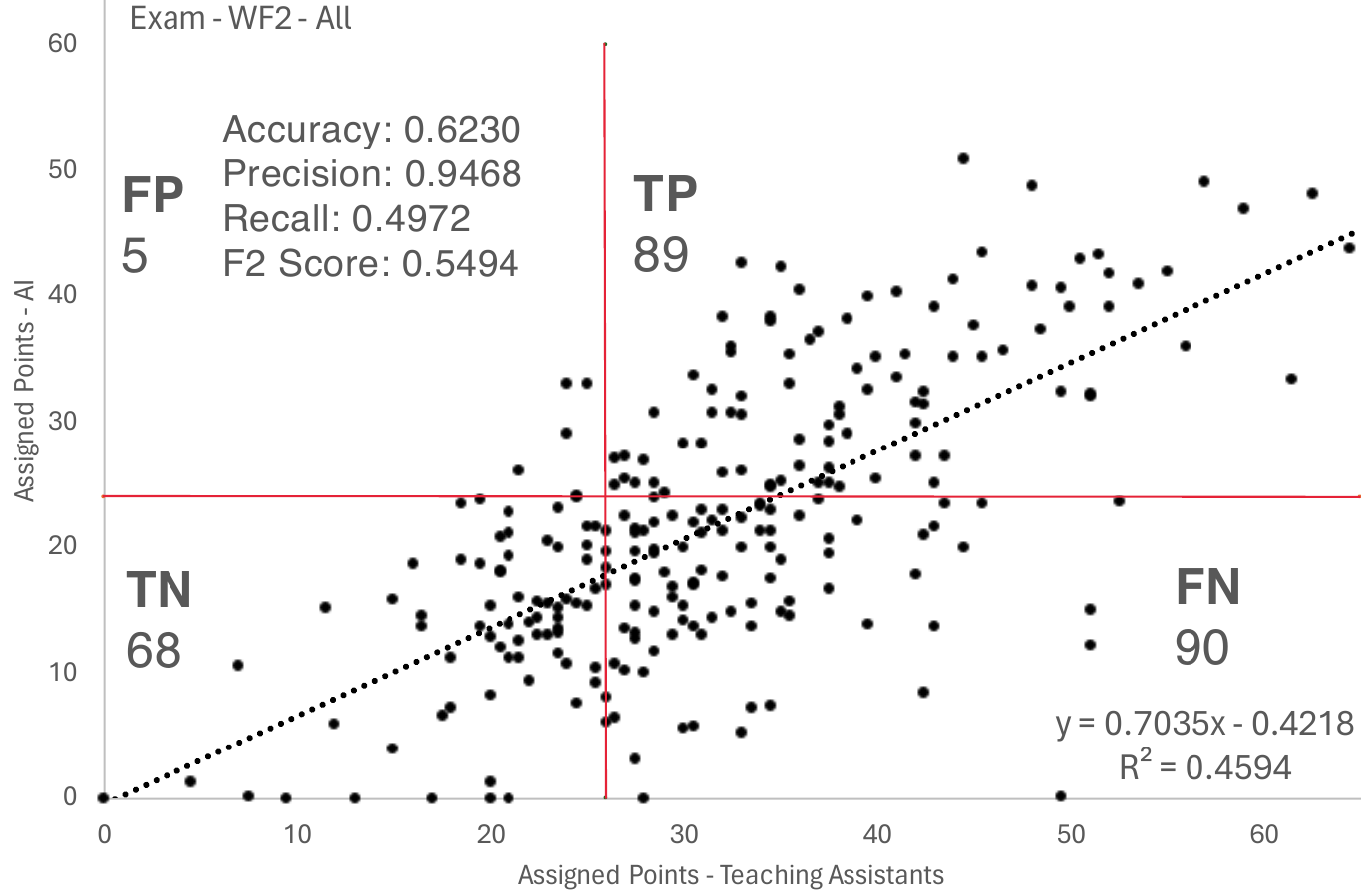}
\includegraphics[width=0.49\textwidth]{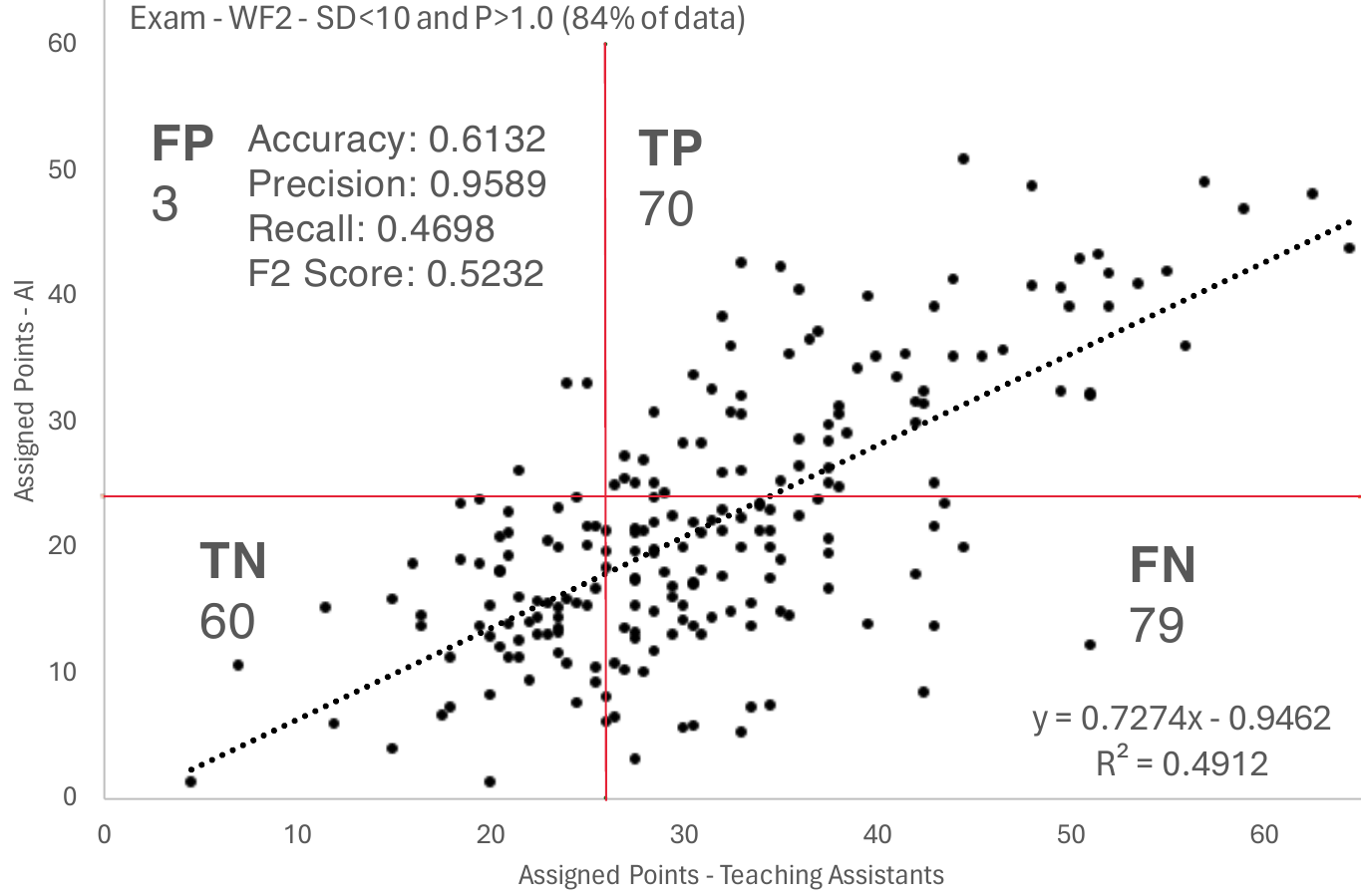}
\end{center}
\caption{Total exam points using WF2. The graphs show the linear interpolations, as well as the contingency table and quality measures for passing the exam.}
\label{fig:totalexam}
\end{figure*} 
\subsection{Total Exam Points and Passing or Failing}
Focussing on WF2, the overall grading performance is less than convincing: the Coefficient of Determination  between the total points determined by the TAs and those determined by AI is $R^2=0.46$. When setting some additional thresholds, namely that the sum of all standard deviations is less than 10 points ($\sum\sigma<10$) to eliminate data points with low confidence, and also eliminating data points with less than one point total, as they are likely the result of a failure of the OCR process, one is left with $84\%$ of the data and  $R^2=0.49$.

Applying the same rule for the determination of the passing threshold as with the TA-score, 50\% of the best AI-score is 25~points. The next gap in the score distribution occurs at 24~points, which would be a passing threshold that could be automatically determined from the AI-grades. One can define the following:
\begin{itemize}
\item True Positive (TP): the AI grading passed a student who also passed with TA grading.
\item True Negative (TN): the AI grading failed a student who also failed with TA grading
\item False Positive (FP): the AI grading passed a student who failed with TA grading
\item False Negative (FN): the AI grading failed a student who passed with TA grading
\end{itemize}
The left panel of Fig.~\ref{fig:totalexam} shows the contingency table (separated by the red lines) for all 252~students, the right panel shows the same for the above outlined restrictions.  Also indicated are the standard measures accuracy, precision, recall, and F2-score:
\begin{eqnarray*}
\text{Accuracy} & = & \frac{\mbox{TP} + \mbox{TN}}{\mbox{TP} + \mbox{TN} + \mbox{FP} + \mbox{FN}} \\
\text{Precision} & = & \frac{\mbox{TP}}{\mbox{TP} + \mbox{FP}} \\
\text{Recall} & = & \frac{\mbox{TP}}{\mbox{TP} + \mbox{FN}} \\
\text{F2-score}& = & \frac{(1 + 2^2) \times \text{Precision} \times \text{Recall}}{2^2 \times \text{Precision} + \text{Recall}}
\end{eqnarray*}
Precision measures the ratio of correctly AI-assigned passing grades to the total number of AI-assigned passing grades, while recall measures the ratio of the correctly AI-assigned passing grades to the total number of TA-assigned passing grades.
The workflow has very high precision, (0.9468 and~0.9589 without and with threshold criteria, respectively), but low recall (0.4972 and~0.4698, respectively). Thus, if the AI passed a student, very likely he or she actually passed the exam, but the AI missed more than half of the actually passing grades. The low recall is also what causes the low accuracy and the low F2-score.

\subsection{Summary and General Observations}
Two of the investigated workflows were hampered by the fact that LLMs are simply not good at bookkeeping: for WF1, it would lose rubric items, for WF3, it could not add numbers. All workflows were hampered by OCR, and none of them dealt well with diagrams like Fig.~\ref{fig:studentgraphdesc}.

Overall, model WF1, going by rubric, seems to be the most reliable workflow. In this study, it was not pursued for the whole exam because GPT-4-32k very frequently could not keep track of all of the fine-grained rubric items, leading to an unjustifiable large number of wasted tokens. A hybrid approach, running the model for each part with a handful of rubric items each might be a promising future approach. While that would entail four or five times more grading rounds (based on the number of problem parts), each one of those would likely return a viable result, as the system can reliably keep track of a handful of items.

\section{Lessons Learned}\label{sec:lessons}
Between pseudonymization, sorting, rotating, marking up, and cleaning of the PDFs, considerable human effort went into the AI-assisted grading workflows, i.e., between the scanning and OCR process. When endeavoring to use the outlined workflows in production, a balance has to be found between the students working in a manner that does not distract from the subject matter and minimizing human effort on the part of the course personnel.
\subsection{Avoid Boxes}
During the OCR process, boxes created a surprising amount of problems: frames on paper that the students brought in or boxes that they drew around answers can lead to the OCR software identifying the box as a figure. Particularly problematic was if students wrote across the lines of a box or literally ``outside the box.'' While boxes can guide the human eye, they are a distraction to OCR software and should be avoided.
\subsection{Straight-line Layout}
An advantage of handwriting mathematical derivations is the ability to easily annotate terms (e.g., an arrow or underbrace, ``this goes to zero,'' or sub-calculations literally on the side). While OCR software is surprisingly good dealing with the two-dimensional character of mathematical expressions (e.g., nested fractions, even within column vectors), unfortunately, they cannot follow less structured graphical layouts, and those free-floating texts or calculations end up out of context within the flow of the larger calculations. Here, students might be asked to please stay within horizontal lines, also avoiding side-by-side multi-column layouts.
\subsection{Use Plain Paper}
The presence of lines and checkers on paper as in Fig.~\ref{fig:prob1hand} is surprisingly well ignored by the OCR processes, but still can cause further complications. A better approach is using plain paper, which leads to the next point:
\subsection{Provide the Paper}
Handwriting the problem derivations using pen-and-paper remains essential, but the bring-your-own-paper policy has created logistics problems, and redacting, sorting, rotating, and labeling the pages required considerable effort. OCR can read typewritten information with very high precision, and this feature can be used to automate several of these operations. 

For exams that expect relatively short answers, that is, just a few lines of derivations, simply provide space on the exam sheet. Instead of answer boxes, leave marker texts according to which the OCR'd text file can be separated, for example, print ``Solution Problem \#2:'' at the beginning of the free space, and ``Problem \#3:'' before the text of the next problem; markers like these do not distract the students but can easily be picked up by the evaluation scripts between the OCR and the grading.

\begin{figure*}
\begin{center}
\includegraphics[width=0.74\textwidth]{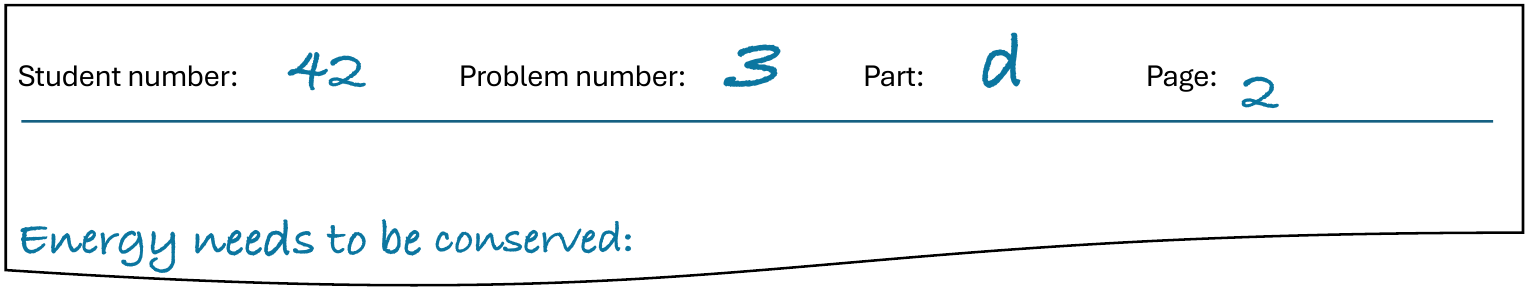}
\end{center}
\caption{Suggested header for instructor-provided exam paper.}
\label{fig:sampleheader}
\end{figure*}

For exams that require longer derivations of unclear length (depending on the approach taken), a simple step would be to provide plain paper that has a normed header with a defined vocabulary. This header might look like the one depicted in Fig.~\ref{fig:sampleheader}. Note that there are no boxes. The terms ``Student number, problem number, part, and page'' will be read by the OCR software with very high precision, and subsequent scripts can easily divide and sort pages according to this pattern (note that it does not matter if the page number is absolute or relative to the problem start or the part start, as the pages would be first sorted by problem, then part, then number). Torn-out pages from notebooks  or oversized sheets can also cause paper jam during the automated scanning.

Students would be instructed to not have more than one problem part on the same page. They would also be asked to use the paper in portrait orientation with the header on top, minimizing the need for later page rotations. Any pattern-matching would need to tolerant enough to deal with students' errors that might be introduced by nervousness.

\subsection{Encourage Students to Write More}
As can be seen when comparing Figs.~\ref{fig:examplehand} and~\ref{fig:sample}, students tend to write much less than experts when solving problems. It is good practice to provide more explanations when deriving the answers to problems, and from a student point of view, it also provides more chances to get partial credit when the overall solution turns out to be wrong. For AI-grading, having more rather than less text and written out formula derivations greatly increases error-tolerance in both the OCR and the grading process. LLMs tend to be verbose in their answers, but they also appear to need the similar verbosity to make reliable statements on similarity; they are, after all, probabilistic language models, not deterministic symbolic algebra systems.

\subsection{Use Pencils and Erasers}
Scribbling out writing is not understood by OCR software; at times, the OCR software provides an interpretation of these expressions that the student wanted ignored, at other times, the model start hallucinating (see Fig.~\ref{fig:oddity}). Permanent markers are oftentimes mandated to avoid tampering with the exams once they are returned to the student, but the first step in this process is scanning. Thus, an electronic copy of the original student work is available, which would not only provide clear evidence of later tampering, but likely prevent this dishonest action in the first place. Students should erase wrong expressions rather than scribbling them out (providing extra pencils and clean, white erasers in the exam room is probably a good idea).

\subsection{Avoid Graphical Problems (for now)}
Our experiments showed that least for the moment, various steps in the workflows fail for graphical problems: the graphs might get ``overlooked,'' and their descriptions might be too vague to allow for a meaningful comparison to sample solutions.

\section{Limitations}
This study is decidedly empirical, investigating different workflows for AI-assisted grading of one thermodynamics using tool currently available (Spring 2024). The results are specific to the exam, the rubric (Table~\ref{tab:grading_rubric}), and the prompts (Figs.~\ref{fig:prompt_wf1}--\ref{fig:prompt_wf3}), so only limited generalizability can be claimed.

\section{Outlook}
The logistics and format of high-stakes exams are hard to adapt according to the lessons in Sect.~\ref{sec:lessons}, and a future study should consider a lower-stakes exam which can be adapted for AI-grading assistance, even if this comes at the expense of authenticity of the data. Based on the findings in this study, a workflow should be investigated that uses a detailed grading rubric, but only applies it to one problem part at a time to reduce the number of failed grading attempts.

As even humans can sometimes only decipher handwritten solutions like Fig.~\ref{fig:examplehand} in context, future experiments with GPT-only OCR processes (like in WF4) might include the problem text in the reading prompt. Preliminary findings indicated that currently, this runs the risk of pushing the system over the token limit, and it can lead to increased hallucinations. In particular, the process runs the risk of the LLM attempting to solve the problem and ``seeing what it expects to see;'' further and more diligent prompt engineering would be needed to ensure that the system avoids OCR errors but does not fix physics errors.This is particularly true since currently when processing a page, it is unknown which problem is being answered, so at the moment, all problem texts would need to be included. It can be expected that future models have higher token limits.

\begin{figure}
\begin{center}
\includegraphics[width=0.9\columnwidth]{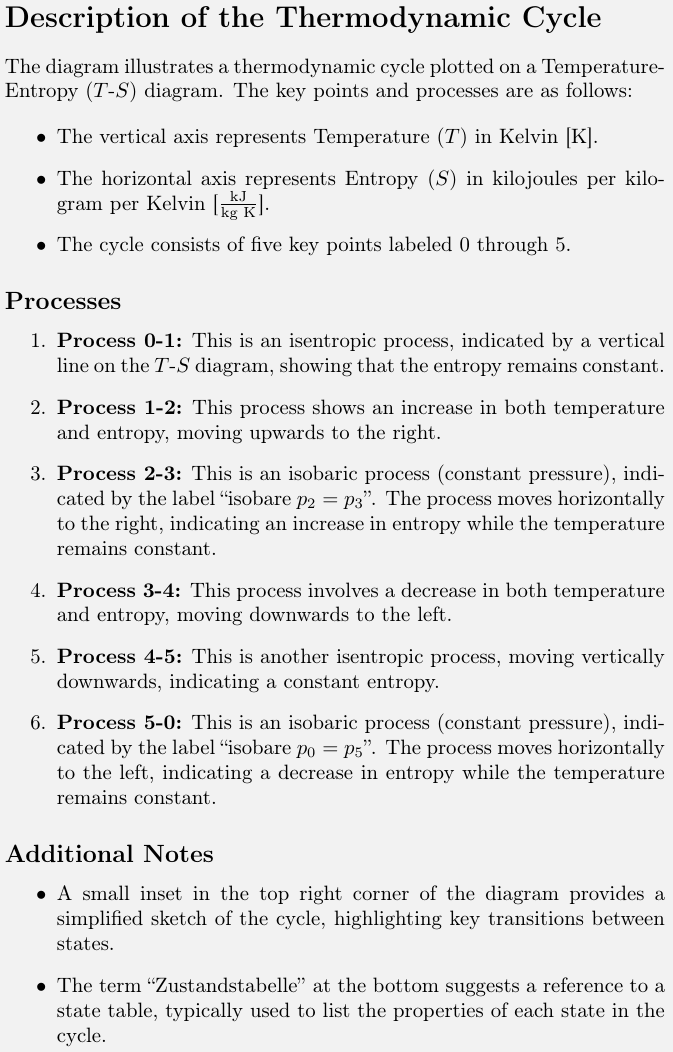}
\end{center}
\caption{GPT-4o description of the graph in Fig.~\ref{fig:studentgraphdesc}}
\label{fig:GPT4oGraph}
\end{figure}

At the time of our study, GPT-4o~\cite{gpt4o} had not been available yet. There is anecdotal evidence that GPT-4o is worse than GPT-4 in mathematical and reasoning task, but better in handwriting recognition. For example, the excerpt in  Fig.~\ref{fig:examplehand} gets interpreted as

\begin{equation*}
    \text{d)} \quad U = u_{f2} + x \left( u_{g} - u_{f2} \right) - m g (u_2 - u_1) + Q = W_v
\end{equation*}

\begin{equation*}
    U \left(0.003 \, c \right) \Rightarrow u_{Fc} = -333.492 \quad u_{Fc} = -0.033
\end{equation*}

\begin{equation*}
    x_{f2} = \frac{U_{n+2} - U_{f2}}{u_g - u_{f2}}
\end{equation*}

which is better than the results in Fig.~\ref{fig:ocr}. Also, GPT-4o shows promise in interpreting graphs; Fig.~\ref{fig:GPT4oGraph} shows the GPT-4o description of the graph in Fig.~\ref{fig:studentgraphdesc}.

Future work will explore a grading workflow that uses GPT-4o for handwriting and graphics interpretation, followed by using GPT-4 to grade each individual problem part with a detailed grading rubric, and summing up points using Python.

\section{Conclusion}
While we found once again that Large Language Models (LLMs) can neither reliably count nor add numbers, they have unique properties that make them useful for assisting in free-form exam grading: they can deal with fuzzy data in a probabilistic manner. 

The biggest source of uncertainty was the conversion of handwriting to a machine-readable format, that is, the Optical Character Recognition which forms the base for all further grading steps. We found that some common sources of error can be addressed by changing the format of the exam sheets, and we had anecdotal evidence that newer multimodal LLMs perform better.

Once the exams are in machine-readable format, in our case LaTeX, we found that the granularity of the grading prompts is essential to both the accuracy and the rate-of-failure of automated grading assistance. A fine-grained grading rubric, applied to a whole problem at a time, leads to frequent bookkeeping errors and failed grading attempts. Grading the whole problems by a handful of parts, using the full sample solution, turned out to be more reliable, but misses some of the nuances and weightings of a rubric. When attempting to grade the whole problem all at once, the model resorted to grading it by part, but then failed to correctly add up the points; essentially, again, a bookkeeping error.

Grading graphical solutions, in our case process diagrams, turned out to have much lower reliability than grading mathematical derivations. When those solutions were not ``overlooked'' in the first place, the descriptions of free-hand drawings were much less reliable than those resulting from careful construction on a computer; crooked lines or scribbles provided too much extraneous information, which the tool was unable to distinguish from the salient information. Also here, newer multimodal models may perform better.

The overall gradings had high precision, that is, they identified exams that met passing criteria with high reliability. However, other performance measures such as accuracy, recall, and F2 were too low for failing students; exams that would lead to failing grades definitely require human validation. Noticeable were also high-confidence zero-point grades, which result from utter OCR failures.

\begin{acknowledgments}
We would like to thank the students who participated in this study. We would also like to thank Fadoua Balabdaoui for statistical consulting, and Alina Yaroshchuk for assistance with exam logistics.
\end{acknowledgments}

\bibliography{thermoexam}% Produces the bibliography via BibTeX.

%apsrev4-2.bst 2019-01-14 (MD) hand-edited version of apsrev4-1.bst
%Control: key (0)
%Control: author (8) initials jnrlst
%Control: editor formatted (1) identically to author
%Control: production of article title (0) allowed
%Control: page (0) single
%Control: year (1) truncated
%Control: production of eprint (0) enabled
\begin{thebibliography}{52}%
\makeatletter
\providecommand \@ifxundefined [1]{%
 \@ifx{#1\undefined}
}%
\providecommand \@ifnum [1]{%
 \ifnum #1\expandafter \@firstoftwo
 \else \expandafter \@secondoftwo
 \fi
}%
\providecommand \@ifx [1]{%
 \ifx #1\expandafter \@firstoftwo
 \else \expandafter \@secondoftwo
 \fi
}%
\providecommand \natexlab [1]{#1}%
\providecommand \enquote  [1]{``#1''}%
\providecommand \bibnamefont  [1]{#1}%
\providecommand \bibfnamefont [1]{#1}%
\providecommand \citenamefont [1]{#1}%
\providecommand \href@noop [0]{\@secondoftwo}%
\providecommand \href [0]{\begingroup \@sanitize@url \@href}%
\providecommand \@href[1]{\@@startlink{#1}\@@href}%
\providecommand \@@href[1]{\endgroup#1\@@endlink}%
\providecommand \@sanitize@url [0]{\catcode `\\12\catcode `\$12\catcode
  `\&12\catcode `\#12\catcode `\^12\catcode `\_12\catcode `\%12\relax}%
\providecommand \@@startlink[1]{}%
\providecommand \@@endlink[0]{}%
\providecommand \url  [0]{\begingroup\@sanitize@url \@url }%
\providecommand \@url [1]{\endgroup\@href {#1}{\urlprefix }}%
\providecommand \urlprefix  [0]{URL }%
\providecommand \Eprint [0]{\href }%
\providecommand \doibase [0]{https://doi.org/}%
\providecommand \selectlanguage [0]{\@gobble}%
\providecommand \bibinfo  [0]{\@secondoftwo}%
\providecommand \bibfield  [0]{\@secondoftwo}%
\providecommand \translation [1]{[#1]}%
\providecommand \BibitemOpen [0]{}%
\providecommand \bibitemStop [0]{}%
\providecommand \bibitemNoStop [0]{.\EOS\space}%
\providecommand \EOS [0]{\spacefactor3000\relax}%
\providecommand \BibitemShut  [1]{\csname bibitem#1\endcsname}%
\let\auto@bib@innerbib\@empty
%</preamble>
\bibitem [{\citenamefont {{OpenAI}}(2024{\natexlab{a}})}]{chatgpt}%
  \BibitemOpen
  \bibfield  {author} {\bibinfo {author} {\bibnamefont {{OpenAI}}},\
  }\href@noop {} {\bibinfo {title} {{ChatGPT}}},\ \bibinfo {howpublished}
  {\url{https://chat.openai.com/}} (\bibinfo {year} {accessed April
  2024}{\natexlab{a}})\BibitemShut {NoStop}%
\bibitem [{\citenamefont {Turing}(1950)}]{turing1950}%
  \BibitemOpen
  \bibfield  {author} {\bibinfo {author} {\bibfnamefont {A.~M.}\ \bibnamefont
  {Turing}},\ }\bibfield  {title} {\bibinfo {title} {Computing machinery and
  intelligence},\ }\href {https://doi.org/10.1093/mind/LIX.236.433} {\bibfield
  {journal} {\bibinfo  {journal} {Mind}\ ,\ \bibinfo {pages} {433}} (\bibinfo
  {year} {1950})}\BibitemShut {NoStop}%
\bibitem [{\citenamefont {{OpenAI}}(2024{\natexlab{b}})}]{gpt4}%
  \BibitemOpen
  \bibfield  {author} {\bibinfo {author} {\bibnamefont {{OpenAI}}},\
  }\href@noop {} {\bibinfo {title} {{ChatGPT}}},\ \bibinfo {howpublished}
  {\url{https://openai.com/research/gpt-4}} (\bibinfo {year} {accessed April
  2024}{\natexlab{b}})\BibitemShut {NoStop}%
\bibitem [{\citenamefont {Kung}\ \emph {et~al.}(2022)\citenamefont {Kung},
  \citenamefont {Cheatham}, \citenamefont {Medinilla}, \citenamefont {ChatGPT},
  \citenamefont {Sillos}, \citenamefont {De~Leon}, \citenamefont {Elepano},
  \citenamefont {Madriaga}, \citenamefont {Aggabao}, \citenamefont
  {Diaz-Candido} \emph {et~al.}}]{kung2022}%
  \BibitemOpen
  \bibfield  {author} {\bibinfo {author} {\bibfnamefont {T.~H.}\ \bibnamefont
  {Kung}}, \bibinfo {author} {\bibfnamefont {M.}~\bibnamefont {Cheatham}},
  \bibinfo {author} {\bibfnamefont {A.}~\bibnamefont {Medinilla}}, \bibinfo
  {author} {\bibnamefont {ChatGPT}}, \bibinfo {author} {\bibfnamefont
  {C.}~\bibnamefont {Sillos}}, \bibinfo {author} {\bibfnamefont
  {L.}~\bibnamefont {De~Leon}}, \bibinfo {author} {\bibfnamefont
  {C.}~\bibnamefont {Elepano}}, \bibinfo {author} {\bibfnamefont
  {M.}~\bibnamefont {Madriaga}}, \bibinfo {author} {\bibfnamefont
  {R.}~\bibnamefont {Aggabao}}, \bibinfo {author} {\bibfnamefont
  {G.}~\bibnamefont {Diaz-Candido}}, \emph {et~al.},\ }\bibfield  {title}
  {\bibinfo {title} {Performance of chatgpt on usmle: Potential for ai-assisted
  medical education using large language models},\ }\href@noop {} {\bibfield
  {journal} {\bibinfo  {journal} {medRxiv}\ ,\ \bibinfo {pages} {2022}}
  (\bibinfo {year} {2022})}\BibitemShut {NoStop}%
\bibitem [{\citenamefont {Achiam}\ \emph {et~al.}(2023)\citenamefont {Achiam},
  \citenamefont {Adler}, \citenamefont {Agarwal}, \citenamefont {Ahmad},
  \citenamefont {Akkaya}, \citenamefont {Aleman}, \citenamefont {Almeida},
  \citenamefont {Altenschmidt}, \citenamefont {Altman}, \citenamefont {Anadkat}
  \emph {et~al.}}]{achiam2023gpt}%
  \BibitemOpen
  \bibfield  {author} {\bibinfo {author} {\bibfnamefont {J.}~\bibnamefont
  {Achiam}}, \bibinfo {author} {\bibfnamefont {S.}~\bibnamefont {Adler}},
  \bibinfo {author} {\bibfnamefont {S.}~\bibnamefont {Agarwal}}, \bibinfo
  {author} {\bibfnamefont {L.}~\bibnamefont {Ahmad}}, \bibinfo {author}
  {\bibfnamefont {I.}~\bibnamefont {Akkaya}}, \bibinfo {author} {\bibfnamefont
  {F.~L.}\ \bibnamefont {Aleman}}, \bibinfo {author} {\bibfnamefont
  {D.}~\bibnamefont {Almeida}}, \bibinfo {author} {\bibfnamefont
  {J.}~\bibnamefont {Altenschmidt}}, \bibinfo {author} {\bibfnamefont
  {S.}~\bibnamefont {Altman}}, \bibinfo {author} {\bibfnamefont
  {S.}~\bibnamefont {Anadkat}}, \emph {et~al.},\ }\bibfield  {title} {\bibinfo
  {title} {{GPT-4} technical report},\ }\href@noop {} {\bibfield  {journal}
  {\bibinfo  {journal} {arXiv preprint arXiv:2303.08774}\ } (\bibinfo {year}
  {2023})}\BibitemShut {NoStop}%
\bibitem [{\citenamefont {{Samantha Murphy Kelly}}(2023)}]{lawexam}%
  \BibitemOpen
  \bibfield  {author} {\bibinfo {author} {\bibnamefont {{Samantha Murphy
  Kelly}}},\ }\href@noop {} {\bibinfo {title} {{ChatGPT} passes exams from law
  and business schools}},\ \bibinfo {howpublished}
  {\url{https://edition.cnn.com/2023/01/26/tech/chatgpt-passes-exams/index.html}}
  (\bibinfo {year} {accessed January 2023})\BibitemShut {NoStop}%
\bibitem [{\citenamefont {Kortemeyer}(2023{\natexlab{a}})}]{kortemeyer23ai}%
  \BibitemOpen
  \bibfield  {author} {\bibinfo {author} {\bibfnamefont {G.}~\bibnamefont
  {Kortemeyer}},\ }\bibfield  {title} {\bibinfo {title} {Could an
  artificial-intelligence agent pass an introductory physics course?},\ }\href
  {https://doi.org/10.1103/PhysRevPhysEducRes.19.010132} {\bibfield  {journal}
  {\bibinfo  {journal} {Phys. Rev. Phys. Educ. Res.}\ }\textbf {\bibinfo
  {volume} {19}},\ \bibinfo {pages} {010132} (\bibinfo {year}
  {2023}{\natexlab{a}})}\BibitemShut {NoStop}%
\bibitem [{\citenamefont {Polverini}\ and\ \citenamefont
  {Gregorcic}(2024)}]{polverini24}%
  \BibitemOpen
  \bibfield  {author} {\bibinfo {author} {\bibfnamefont {G.}~\bibnamefont
  {Polverini}}\ and\ \bibinfo {author} {\bibfnamefont {B.}~\bibnamefont
  {Gregorcic}},\ }\bibfield  {title} {\bibinfo {title} {Performance of chatgpt
  on the test of understanding graphs in kinematics},\ }\href
  {https://doi.org/10.1103/PhysRevPhysEducRes.20.010109} {\bibfield  {journal}
  {\bibinfo  {journal} {Phys. Rev. Phys. Educ. Res.}\ }\textbf {\bibinfo
  {volume} {20}},\ \bibinfo {pages} {010109} (\bibinfo {year}
  {2024})}\BibitemShut {NoStop}%
\bibitem [{\citenamefont {Kortemeyer}\ and\ \citenamefont
  {Bauer}(2024)}]{kortemeyer24cheating}%
  \BibitemOpen
  \bibfield  {author} {\bibinfo {author} {\bibfnamefont {G.}~\bibnamefont
  {Kortemeyer}}\ and\ \bibinfo {author} {\bibfnamefont {W.}~\bibnamefont
  {Bauer}},\ }\bibfield  {title} {\bibinfo {title} {Cheat sites and artificial
  intelligence usage in online introductory physics courses: What is the extent
  and what effect does it have on assessments?},\ }\href
  {https://doi.org/10.1103/PhysRevPhysEducRes.20.010145} {\bibfield  {journal}
  {\bibinfo  {journal} {Phys. Rev. Phys. Educ. Res.}\ }\textbf {\bibinfo
  {volume} {20}},\ \bibinfo {pages} {010145} (\bibinfo {year}
  {2024})}\BibitemShut {NoStop}%
\bibitem [{\citenamefont {Yeadon}\ and\ \citenamefont
  {Hardy}(2024)}]{yeadon2024impact}%
  \BibitemOpen
  \bibfield  {author} {\bibinfo {author} {\bibfnamefont {W.}~\bibnamefont
  {Yeadon}}\ and\ \bibinfo {author} {\bibfnamefont {T.}~\bibnamefont {Hardy}},\
  }\bibfield  {title} {\bibinfo {title} {The impact of {AI} in physics
  education: a comprehensive review from {GCSE} to university levels},\
  }\href@noop {} {\bibfield  {journal} {\bibinfo  {journal} {Physics
  Education}\ }\textbf {\bibinfo {volume} {59}},\ \bibinfo {pages} {025010}
  (\bibinfo {year} {2024})}\BibitemShut {NoStop}%
\bibitem [{\citenamefont {Sperling}\ and\ \citenamefont
  {Lincoln}(2024)}]{sperling2024artificial}%
  \BibitemOpen
  \bibfield  {author} {\bibinfo {author} {\bibfnamefont {A.}~\bibnamefont
  {Sperling}}\ and\ \bibinfo {author} {\bibfnamefont {J.}~\bibnamefont
  {Lincoln}},\ }\bibfield  {title} {\bibinfo {title} {Artificial intelligence
  and high school physics},\ }\href@noop {} {\bibfield  {journal} {\bibinfo
  {journal} {The Physics Teacher}\ }\textbf {\bibinfo {volume} {62}},\ \bibinfo
  {pages} {314} (\bibinfo {year} {2024})}\BibitemShut {NoStop}%
\bibitem [{\citenamefont {K\"uchemann}\ \emph {et~al.}(2023)\citenamefont
  {K\"uchemann}, \citenamefont {Steinert}, \citenamefont {Revenga},
  \citenamefont {Schweinberger}, \citenamefont {Dinc}, \citenamefont {Avila},\
  and\ \citenamefont {Kuhn}}]{kuchemann23}%
  \BibitemOpen
  \bibfield  {author} {\bibinfo {author} {\bibfnamefont {S.}~\bibnamefont
  {K\"uchemann}}, \bibinfo {author} {\bibfnamefont {S.}~\bibnamefont
  {Steinert}}, \bibinfo {author} {\bibfnamefont {N.}~\bibnamefont {Revenga}},
  \bibinfo {author} {\bibfnamefont {M.}~\bibnamefont {Schweinberger}}, \bibinfo
  {author} {\bibfnamefont {Y.}~\bibnamefont {Dinc}}, \bibinfo {author}
  {\bibfnamefont {K.~E.}\ \bibnamefont {Avila}},\ and\ \bibinfo {author}
  {\bibfnamefont {J.}~\bibnamefont {Kuhn}},\ }\bibfield  {title} {\bibinfo
  {title} {Can {ChatGPT} support prospective teachers in physics task
  development?},\ }\href {https://doi.org/10.1103/PhysRevPhysEducRes.19.020128}
  {\bibfield  {journal} {\bibinfo  {journal} {Phys. Rev. Phys. Educ. Res.}\
  }\textbf {\bibinfo {volume} {19}},\ \bibinfo {pages} {020128} (\bibinfo
  {year} {2023})}\BibitemShut {NoStop}%
\bibitem [{\citenamefont
  {Kortemeyer}(2023{\natexlab{b}})}]{kortemeyer2023using}%
  \BibitemOpen
  \bibfield  {author} {\bibinfo {author} {\bibfnamefont {G.}~\bibnamefont
  {Kortemeyer}},\ }\bibfield  {title} {\bibinfo {title} {Using
  artificial-intelligence tools to make {LaTeX} content accessible to blind
  readers},\ }\href@noop {} {\bibfield  {journal} {\bibinfo  {journal}
  {TUGboat}\ }\textbf {\bibinfo {volume} {44}},\ \bibinfo {pages} {390}
  (\bibinfo {year} {2023}{\natexlab{b}})}\BibitemShut {NoStop}%
\bibitem [{\citenamefont {Tschisgale}\ \emph {et~al.}(2023)\citenamefont
  {Tschisgale}, \citenamefont {Wulff},\ and\ \citenamefont
  {Kubsch}}]{tschisgale2023integrating}%
  \BibitemOpen
  \bibfield  {author} {\bibinfo {author} {\bibfnamefont {P.}~\bibnamefont
  {Tschisgale}}, \bibinfo {author} {\bibfnamefont {P.}~\bibnamefont {Wulff}},\
  and\ \bibinfo {author} {\bibfnamefont {M.}~\bibnamefont {Kubsch}},\
  }\bibfield  {title} {\bibinfo {title} {Integrating artificial
  intelligence-based methods into qualitative research in physics education
  research: A case for computational grounded theory},\ }\href@noop {}
  {\bibfield  {journal} {\bibinfo  {journal} {Physical Review Physics Education
  Research}\ }\textbf {\bibinfo {volume} {19}},\ \bibinfo {pages} {020123}
  (\bibinfo {year} {2023})}\BibitemShut {NoStop}%
\bibitem [{\citenamefont {Kieser}\ \emph {et~al.}(2023)\citenamefont {Kieser},
  \citenamefont {Wulff}, \citenamefont {Kuhn},\ and\ \citenamefont
  {K\"uchemann}}]{kieser23}%
  \BibitemOpen
  \bibfield  {author} {\bibinfo {author} {\bibfnamefont {F.}~\bibnamefont
  {Kieser}}, \bibinfo {author} {\bibfnamefont {P.}~\bibnamefont {Wulff}},
  \bibinfo {author} {\bibfnamefont {J.}~\bibnamefont {Kuhn}},\ and\ \bibinfo
  {author} {\bibfnamefont {S.}~\bibnamefont {K\"uchemann}},\ }\bibfield
  {title} {\bibinfo {title} {Educational data augmentation in physics education
  research using {ChatGPT}},\ }\href
  {https://doi.org/10.1103/PhysRevPhysEducRes.19.020150} {\bibfield  {journal}
  {\bibinfo  {journal} {Phys. Rev. Phys. Educ. Res.}\ }\textbf {\bibinfo
  {volume} {19}},\ \bibinfo {pages} {020150} (\bibinfo {year}
  {2023})}\BibitemShut {NoStop}%
\bibitem [{\citenamefont {Reif}\ \emph {et~al.}(1976)\citenamefont {Reif},
  \citenamefont {Larkin},\ and\ \citenamefont {Brackett}}]{reif1976}%
  \BibitemOpen
  \bibfield  {author} {\bibinfo {author} {\bibfnamefont {F.}~\bibnamefont
  {Reif}}, \bibinfo {author} {\bibfnamefont {J.~H.}\ \bibnamefont {Larkin}},\
  and\ \bibinfo {author} {\bibfnamefont {G.~C.}\ \bibnamefont {Brackett}},\
  }\bibfield  {title} {\bibinfo {title} {Teaching general learning and
  problem-solving skills},\ }\href@noop {} {\bibfield  {journal} {\bibinfo
  {journal} {American Journal of Physics}\ }\textbf {\bibinfo {volume} {44}},\
  \bibinfo {pages} {212} (\bibinfo {year} {1976})}\BibitemShut {NoStop}%
\bibitem [{\citenamefont {Reif}(1995)}]{reif1995}%
  \BibitemOpen
  \bibfield  {author} {\bibinfo {author} {\bibfnamefont {F.}~\bibnamefont
  {Reif}},\ }\bibfield  {title} {\bibinfo {title} {Millikan lecture 1994:
  Understanding and teaching important scientific thought processes},\
  }\href@noop {} {\bibfield  {journal} {\bibinfo  {journal} {American Journal
  of Physics}\ }\textbf {\bibinfo {volume} {63}},\ \bibinfo {pages} {17}
  (\bibinfo {year} {1995})}\BibitemShut {NoStop}%
\bibitem [{\citenamefont {Hsu}\ \emph {et~al.}(2004)\citenamefont {Hsu},
  \citenamefont {Brewe}, \citenamefont {Foster},\ and\ \citenamefont
  {Harper}}]{hsu2004}%
  \BibitemOpen
  \bibfield  {author} {\bibinfo {author} {\bibfnamefont {L.}~\bibnamefont
  {Hsu}}, \bibinfo {author} {\bibfnamefont {E.}~\bibnamefont {Brewe}}, \bibinfo
  {author} {\bibfnamefont {T.~M.}\ \bibnamefont {Foster}},\ and\ \bibinfo
  {author} {\bibfnamefont {K.~A.}\ \bibnamefont {Harper}},\ }\bibfield  {title}
  {\bibinfo {title} {Resource letter rps-1: Research in problem solving},\
  }\href@noop {} {\bibfield  {journal} {\bibinfo  {journal} {American journal
  of physics}\ }\textbf {\bibinfo {volume} {72}},\ \bibinfo {pages} {1147}
  (\bibinfo {year} {2004})}\BibitemShut {NoStop}%
\bibitem [{\citenamefont {Hattie}(2008)}]{hattie2008}%
  \BibitemOpen
  \bibfield  {author} {\bibinfo {author} {\bibfnamefont {J.}~\bibnamefont
  {Hattie}},\ }\href@noop {} {\emph {\bibinfo {title} {Visible learning: A
  synthesis of over 800 meta-analyses relating to achievement}}}\ (\bibinfo
  {publisher} {routledge},\ \bibinfo {year} {2008})\BibitemShut {NoStop}%
\bibitem [{\citenamefont {Al-Salmani}\ \emph {et~al.}(2023)\citenamefont
  {Al-Salmani}, \citenamefont {Johnson},\ and\ \citenamefont
  {Thacker}}]{alsalmani23}%
  \BibitemOpen
  \bibfield  {author} {\bibinfo {author} {\bibfnamefont {F.}~\bibnamefont
  {Al-Salmani}}, \bibinfo {author} {\bibfnamefont {J.}~\bibnamefont
  {Johnson}},\ and\ \bibinfo {author} {\bibfnamefont {B.}~\bibnamefont
  {Thacker}},\ }\bibfield  {title} {\bibinfo {title} {Assessing thinking skills
  in free-response exam problems: Pandemic online and in-person},\ }\href
  {https://doi.org/10.1103/PhysRevPhysEducRes.19.010131} {\bibfield  {journal}
  {\bibinfo  {journal} {Phys. Rev. Phys. Educ. Res.}\ }\textbf {\bibinfo
  {volume} {19}},\ \bibinfo {pages} {010131} (\bibinfo {year}
  {2023})}\BibitemShut {NoStop}%
\bibitem [{\citenamefont {Kashy}\ \emph {et~al.}(2001)\citenamefont {Kashy},
  \citenamefont {Albertelli}, \citenamefont {Ashkenazi}, \citenamefont {Kashy},
  \citenamefont {Ng},\ and\ \citenamefont {Thoennessen}}]{kashyd01}%
  \BibitemOpen
  \bibfield  {author} {\bibinfo {author} {\bibfnamefont {D.~A.}\ \bibnamefont
  {Kashy}}, \bibinfo {author} {\bibfnamefont {G.}~\bibnamefont {Albertelli}},
  \bibinfo {author} {\bibfnamefont {G.}~\bibnamefont {Ashkenazi}}, \bibinfo
  {author} {\bibfnamefont {E.}~\bibnamefont {Kashy}}, \bibinfo {author}
  {\bibfnamefont {H.-K.}\ \bibnamefont {Ng}},\ and\ \bibinfo {author}
  {\bibfnamefont {M.}~\bibnamefont {Thoennessen}},\ }\bibfield  {title}
  {\bibinfo {title} {Individualized interactive exercises: a promising role for
  network technology},\ }in\ \href@noop {} {\emph {\bibinfo {booktitle} {Proc.
  Frontiers in Education}}},\ Vol.~\bibinfo {volume} {31}\ (\bibinfo {year}
  {2001})\ pp.\ \bibinfo {pages} {1073--1078}\BibitemShut {NoStop}%
\bibitem [{\citenamefont {Kortemeyer}\ \emph {et~al.}(2008)\citenamefont
  {Kortemeyer}, \citenamefont {Kashy}, \citenamefont {Benenson},\ and\
  \citenamefont {Bauer}}]{kortemeyer08}%
  \BibitemOpen
  \bibfield  {author} {\bibinfo {author} {\bibfnamefont {G.}~\bibnamefont
  {Kortemeyer}}, \bibinfo {author} {\bibfnamefont {E.}~\bibnamefont {Kashy}},
  \bibinfo {author} {\bibfnamefont {W.}~\bibnamefont {Benenson}},\ and\
  \bibinfo {author} {\bibfnamefont {W.}~\bibnamefont {Bauer}},\ }\bibfield
  {title} {\bibinfo {title} {Experiences using the open-source learning content
  management and assessment system {LON-CAPA} in introductory physics
  courses},\ }\href@noop {} {\bibfield  {journal} {\bibinfo  {journal} {Am. J.
  Phys}\ }\textbf {\bibinfo {volume} {76}},\ \bibinfo {pages} {438} (\bibinfo
  {year} {2008})}\BibitemShut {NoStop}%
\bibitem [{\citenamefont {Risley}(2001)}]{risley2001}%
  \BibitemOpen
  \bibfield  {author} {\bibinfo {author} {\bibfnamefont {J.}~\bibnamefont
  {Risley}},\ }\bibfield  {title} {\bibinfo {title} {Motivating students to
  learn physics using an online homework system},\ }\href@noop {} {\bibfield
  {journal} {\bibinfo  {journal} {Newsletter of the APS Forum on Education}\ ,\
  \bibinfo {pages} {3}} (\bibinfo {year} {2001})}\BibitemShut {NoStop}%
\bibitem [{\citenamefont {Stelzer}\ and\ \citenamefont
  {Gladding}(2001)}]{stelzer2001}%
  \BibitemOpen
  \bibfield  {author} {\bibinfo {author} {\bibfnamefont {T.}~\bibnamefont
  {Stelzer}}\ and\ \bibinfo {author} {\bibfnamefont {G.}~\bibnamefont
  {Gladding}},\ }\bibfield  {title} {\bibinfo {title} {The evolution of
  web-based activities in physics at illinois},\ }\href@noop {} {\bibfield
  {journal} {\bibinfo  {journal} {Newsletter of the APS Forum on Education}\ ,\
  \bibinfo {pages} {7}} (\bibinfo {year} {2001})}\BibitemShut {NoStop}%
\bibitem [{\citenamefont {Dufresne}\ \emph {et~al.}(2002)\citenamefont
  {Dufresne}, \citenamefont {Hart}, \citenamefont {Mestre},\ and\ \citenamefont
  {Rath}}]{dufresne02}%
  \BibitemOpen
  \bibfield  {author} {\bibinfo {author} {\bibfnamefont {R.~J.}\ \bibnamefont
  {Dufresne}}, \bibinfo {author} {\bibfnamefont {D.}~\bibnamefont {Hart}},
  \bibinfo {author} {\bibfnamefont {J.~P.}\ \bibnamefont {Mestre}},\ and\
  \bibinfo {author} {\bibfnamefont {K.}~\bibnamefont {Rath}},\ }\bibfield
  {title} {\bibinfo {title} {The effect of web-based homework on test
  performance in large enrollment introductory physics courses},\ }\href@noop
  {} {\bibfield  {journal} {\bibinfo  {journal} {Journal of Computers in
  Mathematics and Science Teaching}\ }\textbf {\bibinfo {volume} {21}},\
  \bibinfo {pages} {229} (\bibinfo {year} {2002})}\BibitemShut {NoStop}%
\bibitem [{\citenamefont {Fredericks}(2007)}]{fredericks2007}%
  \BibitemOpen
  \bibfield  {author} {\bibinfo {author} {\bibfnamefont {C.}~\bibnamefont
  {Fredericks}},\ }\emph {\bibinfo {title} {Patterns of Behavior in Online
  Homework for Introductory Physics}},\ \href@noop {} {Ph.D. thesis},\ \bibinfo
   {school} {University of Massachusetts} (\bibinfo {year} {2007})\BibitemShut
  {NoStop}%
\bibitem [{\citenamefont {Richards-Babb}\ \emph {et~al.}(2011)\citenamefont
  {Richards-Babb}, \citenamefont {Drelick}, \citenamefont {Henry},\ and\
  \citenamefont {Robertson-Honecker}}]{richards2011}%
  \BibitemOpen
  \bibfield  {author} {\bibinfo {author} {\bibfnamefont {M.}~\bibnamefont
  {Richards-Babb}}, \bibinfo {author} {\bibfnamefont {J.}~\bibnamefont
  {Drelick}}, \bibinfo {author} {\bibfnamefont {Z.}~\bibnamefont {Henry}},\
  and\ \bibinfo {author} {\bibfnamefont {J.}~\bibnamefont
  {Robertson-Honecker}},\ }\bibfield  {title} {\bibinfo {title} {Online
  homework, help or hindrance? what students think and how they perform},\
  }\href@noop {} {\bibfield  {journal} {\bibinfo  {journal} {Journal of College
  Science Teaching}\ }\textbf {\bibinfo {volume} {40}},\ \bibinfo {pages} {81}
  (\bibinfo {year} {2011})}\BibitemShut {NoStop}%
\bibitem [{\citenamefont {Perdian}(2013)}]{perdian2013}%
  \BibitemOpen
  \bibfield  {author} {\bibinfo {author} {\bibfnamefont {D.~C.}\ \bibnamefont
  {Perdian}},\ }\bibfield  {title} {\bibinfo {title} {Early identification of
  student performance and effort using an online homework system: A pilot
  study},\ }\href {https://doi.org/10.1007/s10956-012-9423-7} {\bibfield
  {journal} {\bibinfo  {journal} {Journal of Science Education and Technology}\
  }\textbf {\bibinfo {volume} {22}},\ \bibinfo {pages} {697} (\bibinfo {year}
  {2013})}\BibitemShut {NoStop}%
\bibitem [{\citenamefont {Docktor}\ \emph {et~al.}(2016)\citenamefont
  {Docktor}, \citenamefont {Dornfeld}, \citenamefont {Frodermann},
  \citenamefont {Heller}, \citenamefont {Hsu}, \citenamefont {Jackson},
  \citenamefont {Mason}, \citenamefont {Ryan},\ and\ \citenamefont
  {Yang}}]{docktor2016}%
  \BibitemOpen
  \bibfield  {author} {\bibinfo {author} {\bibfnamefont {J.~L.}\ \bibnamefont
  {Docktor}}, \bibinfo {author} {\bibfnamefont {J.}~\bibnamefont {Dornfeld}},
  \bibinfo {author} {\bibfnamefont {E.}~\bibnamefont {Frodermann}}, \bibinfo
  {author} {\bibfnamefont {K.}~\bibnamefont {Heller}}, \bibinfo {author}
  {\bibfnamefont {L.}~\bibnamefont {Hsu}}, \bibinfo {author} {\bibfnamefont
  {K.~A.}\ \bibnamefont {Jackson}}, \bibinfo {author} {\bibfnamefont
  {A.}~\bibnamefont {Mason}}, \bibinfo {author} {\bibfnamefont {Q.~X.}\
  \bibnamefont {Ryan}},\ and\ \bibinfo {author} {\bibfnamefont
  {J.}~\bibnamefont {Yang}},\ }\bibfield  {title} {\bibinfo {title} {Assessing
  student written problem solutions: A problem-solving rubric with application
  to introductory physics},\ }\href@noop {} {\bibfield  {journal} {\bibinfo
  {journal} {Physical review physics education research}\ }\textbf {\bibinfo
  {volume} {12}},\ \bibinfo {pages} {010130} (\bibinfo {year}
  {2016})}\BibitemShut {NoStop}%
\bibitem [{\citenamefont {Burkholder}\ \emph {et~al.}(2020)\citenamefont
  {Burkholder}, \citenamefont {Miles}, \citenamefont {Layden}, \citenamefont
  {Wang}, \citenamefont {Fritz},\ and\ \citenamefont
  {Wieman}}]{burkholder2020}%
  \BibitemOpen
  \bibfield  {author} {\bibinfo {author} {\bibfnamefont {E.}~\bibnamefont
  {Burkholder}}, \bibinfo {author} {\bibfnamefont {J.}~\bibnamefont {Miles}},
  \bibinfo {author} {\bibfnamefont {T.}~\bibnamefont {Layden}}, \bibinfo
  {author} {\bibfnamefont {K.}~\bibnamefont {Wang}}, \bibinfo {author}
  {\bibfnamefont {A.}~\bibnamefont {Fritz}},\ and\ \bibinfo {author}
  {\bibfnamefont {C.}~\bibnamefont {Wieman}},\ }\bibfield  {title} {\bibinfo
  {title} {Template for teaching and assessment of problem solving in
  introductory physics},\ }\href@noop {} {\bibfield  {journal} {\bibinfo
  {journal} {Physical Review Physics Education Research}\ }\textbf {\bibinfo
  {volume} {16}},\ \bibinfo {pages} {010123} (\bibinfo {year}
  {2020})}\BibitemShut {NoStop}%
\bibitem [{\citenamefont {Wan}\ and\ \citenamefont {Chen}(2024)}]{wan24}%
  \BibitemOpen
  \bibfield  {author} {\bibinfo {author} {\bibfnamefont {T.}~\bibnamefont
  {Wan}}\ and\ \bibinfo {author} {\bibfnamefont {Z.}~\bibnamefont {Chen}},\
  }\bibfield  {title} {\bibinfo {title} {Exploring generative {AI} assisted
  feedback writing for students' written responses to a physics conceptual
  question with prompt engineering and few-shot learning},\ }\href
  {https://doi.org/10.1103/PhysRevPhysEducRes.20.010152} {\bibfield  {journal}
  {\bibinfo  {journal} {Phys. Rev. Phys. Educ. Res.}\ }\textbf {\bibinfo
  {volume} {20}},\ \bibinfo {pages} {010152} (\bibinfo {year}
  {2024})}\BibitemShut {NoStop}%
\bibitem [{\citenamefont {Wilson}\ \emph {et~al.}(2022)\citenamefont {Wilson},
  \citenamefont {Pollard}, \citenamefont {Aiken}, \citenamefont {Caballero},\
  and\ \citenamefont {Lewandowski}}]{wilson22}%
  \BibitemOpen
  \bibfield  {author} {\bibinfo {author} {\bibfnamefont {J.}~\bibnamefont
  {Wilson}}, \bibinfo {author} {\bibfnamefont {B.}~\bibnamefont {Pollard}},
  \bibinfo {author} {\bibfnamefont {J.~M.}\ \bibnamefont {Aiken}}, \bibinfo
  {author} {\bibfnamefont {M.~D.}\ \bibnamefont {Caballero}},\ and\ \bibinfo
  {author} {\bibfnamefont {H.~J.}\ \bibnamefont {Lewandowski}},\ }\bibfield
  {title} {\bibinfo {title} {Classification of open-ended responses to a
  research-based assessment using natural language processing},\ }\href
  {https://doi.org/10.1103/PhysRevPhysEducRes.18.010141} {\bibfield  {journal}
  {\bibinfo  {journal} {Phys. Rev. Phys. Educ. Res.}\ }\textbf {\bibinfo
  {volume} {18}},\ \bibinfo {pages} {010141} (\bibinfo {year}
  {2022})}\BibitemShut {NoStop}%
\bibitem [{\citenamefont {Mitros}\ \emph {et~al.}(2013)\citenamefont {Mitros},
  \citenamefont {Paruchuri}, \citenamefont {Rogosic},\ and\ \citenamefont
  {Huang}}]{mitros2013}%
  \BibitemOpen
  \bibfield  {author} {\bibinfo {author} {\bibfnamefont {P.}~\bibnamefont
  {Mitros}}, \bibinfo {author} {\bibfnamefont {V.}~\bibnamefont {Paruchuri}},
  \bibinfo {author} {\bibfnamefont {J.}~\bibnamefont {Rogosic}},\ and\ \bibinfo
  {author} {\bibfnamefont {D.}~\bibnamefont {Huang}},\ }\bibfield  {title}
  {\bibinfo {title} {An integrated framework for the grading of freeform
  responses},\ }in\ \href@noop {} {\emph {\bibinfo {booktitle} {The Sixth
  Conference of MIT's Learning International Networks Consortium}}}\ (\bibinfo
  {year} {2013})\BibitemShut {NoStop}%
\bibitem [{\citenamefont {Dzikovska}\ \emph {et~al.}(2013)\citenamefont
  {Dzikovska}, \citenamefont {Nielsen}, \citenamefont {Brew}, \citenamefont
  {Leacock}, \citenamefont {Giampiccolo}, \citenamefont {Bentivogli},
  \citenamefont {Clark}, \citenamefont {Dagan},\ and\ \citenamefont
  {Dang}}]{dzikovska2013}%
  \BibitemOpen
  \bibfield  {author} {\bibinfo {author} {\bibfnamefont {M.~O.}\ \bibnamefont
  {Dzikovska}}, \bibinfo {author} {\bibfnamefont {R.}~\bibnamefont {Nielsen}},
  \bibinfo {author} {\bibfnamefont {C.}~\bibnamefont {Brew}}, \bibinfo {author}
  {\bibfnamefont {C.}~\bibnamefont {Leacock}}, \bibinfo {author} {\bibfnamefont
  {D.}~\bibnamefont {Giampiccolo}}, \bibinfo {author} {\bibfnamefont
  {L.}~\bibnamefont {Bentivogli}}, \bibinfo {author} {\bibfnamefont
  {P.}~\bibnamefont {Clark}}, \bibinfo {author} {\bibfnamefont
  {I.}~\bibnamefont {Dagan}},\ and\ \bibinfo {author} {\bibfnamefont {H.~T.}\
  \bibnamefont {Dang}},\ }\bibfield  {title} {\bibinfo {title} {Semeval-2013
  task 7: The joint student response analysis and 8th recognizing textual
  entailment challenge},\ }in\ \href@noop {} {\emph {\bibinfo {booktitle}
  {Second Joint Conference on Lexical and Computational Semantics (* SEM),
  Volume 2: Proceedings of the Seventh International Workshop on Semantic
  Evaluation (SemEval 2013)}}}\ (\bibinfo {year} {2013})\ pp.\ \bibinfo {pages}
  {263--274}\BibitemShut {NoStop}%
\bibitem [{\citenamefont {Burrows}\ \emph {et~al.}(2015)\citenamefont
  {Burrows}, \citenamefont {Gurevych},\ and\ \citenamefont
  {Stein}}]{burrows2015}%
  \BibitemOpen
  \bibfield  {author} {\bibinfo {author} {\bibfnamefont {S.}~\bibnamefont
  {Burrows}}, \bibinfo {author} {\bibfnamefont {I.}~\bibnamefont {Gurevych}},\
  and\ \bibinfo {author} {\bibfnamefont {B.}~\bibnamefont {Stein}},\ }\bibfield
   {title} {\bibinfo {title} {The eras and trends of automatic short answer
  grading},\ }\href@noop {} {\bibfield  {journal} {\bibinfo  {journal}
  {International journal of artificial intelligence in education}\ }\textbf
  {\bibinfo {volume} {25}},\ \bibinfo {pages} {60} (\bibinfo {year}
  {2015})}\BibitemShut {NoStop}%
\bibitem [{\citenamefont {Sung}\ \emph {et~al.}(2019)\citenamefont {Sung},
  \citenamefont {Dhamecha}, \citenamefont {Saha}, \citenamefont {Ma},
  \citenamefont {Reddy},\ and\ \citenamefont {Arora}}]{sung2019}%
  \BibitemOpen
  \bibfield  {author} {\bibinfo {author} {\bibfnamefont {C.}~\bibnamefont
  {Sung}}, \bibinfo {author} {\bibfnamefont {T.}~\bibnamefont {Dhamecha}},
  \bibinfo {author} {\bibfnamefont {S.}~\bibnamefont {Saha}}, \bibinfo {author}
  {\bibfnamefont {T.}~\bibnamefont {Ma}}, \bibinfo {author} {\bibfnamefont
  {V.}~\bibnamefont {Reddy}},\ and\ \bibinfo {author} {\bibfnamefont
  {R.}~\bibnamefont {Arora}},\ }\bibfield  {title} {\bibinfo {title}
  {Pre-training {BERT} on domain resources for short answer grading},\ }in\
  \href@noop {} {\emph {\bibinfo {booktitle} {Proceedings of the 2019
  Conference on Empirical Methods in Natural Language Processing and the 9th
  International Joint Conference on Natural Language Processing
  (EMNLP-IJCNLP)}}}\ (\bibinfo {year} {2019})\ pp.\ \bibinfo {pages}
  {6071--6075}\BibitemShut {NoStop}%
\bibitem [{\citenamefont {Azad}\ \emph {et~al.}(2020)\citenamefont {Azad},
  \citenamefont {Chen}, \citenamefont {Fowler}, \citenamefont {West},\ and\
  \citenamefont {Zilles}}]{azad2020}%
  \BibitemOpen
  \bibfield  {author} {\bibinfo {author} {\bibfnamefont {S.}~\bibnamefont
  {Azad}}, \bibinfo {author} {\bibfnamefont {B.}~\bibnamefont {Chen}}, \bibinfo
  {author} {\bibfnamefont {M.}~\bibnamefont {Fowler}}, \bibinfo {author}
  {\bibfnamefont {M.}~\bibnamefont {West}},\ and\ \bibinfo {author}
  {\bibfnamefont {C.}~\bibnamefont {Zilles}},\ }\bibfield  {title} {\bibinfo
  {title} {Strategies for deploying unreliable ai graders in high-transparency
  high-stakes exams},\ }in\ \href@noop {} {\emph {\bibinfo {booktitle}
  {Artificial Intelligence in Education: 21st International Conference, AIED
  2020, Ifrane, Morocco, July 6--10, 2020, Proceedings, Part I 21}}}\ (\bibinfo
  {organization} {Springer},\ \bibinfo {year} {2020})\ pp.\ \bibinfo {pages}
  {16--28}\BibitemShut {NoStop}%
\bibitem [{\citenamefont {Fowler}\ \emph {et~al.}(2021)\citenamefont {Fowler},
  \citenamefont {Chen}, \citenamefont {Azad}, \citenamefont {West},\ and\
  \citenamefont {Zilles}}]{fowler2021}%
  \BibitemOpen
  \bibfield  {author} {\bibinfo {author} {\bibfnamefont {M.}~\bibnamefont
  {Fowler}}, \bibinfo {author} {\bibfnamefont {B.}~\bibnamefont {Chen}},
  \bibinfo {author} {\bibfnamefont {S.}~\bibnamefont {Azad}}, \bibinfo {author}
  {\bibfnamefont {M.}~\bibnamefont {West}},\ and\ \bibinfo {author}
  {\bibfnamefont {C.}~\bibnamefont {Zilles}},\ }\bibfield  {title} {\bibinfo
  {title} {Autograding" explain in plain english" questions using nlp},\ }in\
  \href@noop {} {\emph {\bibinfo {booktitle} {Proceedings of the 52nd ACM
  Technical Symposium on Computer Science Education}}}\ (\bibinfo {year}
  {2021})\ pp.\ \bibinfo {pages} {1163--1169}\BibitemShut {NoStop}%
\bibitem [{\citenamefont {Stanyon}\ \emph {et~al.}(2022)\citenamefont
  {Stanyon}, \citenamefont {Martello}, \citenamefont {Kainth},\ and\
  \citenamefont {Wilkin}}]{stanyon2022}%
  \BibitemOpen
  \bibfield  {author} {\bibinfo {author} {\bibfnamefont {R.}~\bibnamefont
  {Stanyon}}, \bibinfo {author} {\bibfnamefont {E.}~\bibnamefont {Martello}},
  \bibinfo {author} {\bibfnamefont {M.}~\bibnamefont {Kainth}},\ and\ \bibinfo
  {author} {\bibfnamefont {N.~K.}\ \bibnamefont {Wilkin}},\ }\bibfield  {title}
  {\bibinfo {title} {Demo of graide: Ai powered assistive grading engine},\
  }in\ \href@noop {} {\emph {\bibinfo {booktitle} {Proceedings of the Ninth ACM
  Conference on Learning@ Scale}}}\ (\bibinfo {year} {2022})\ pp.\ \bibinfo
  {pages} {466--468}\BibitemShut {NoStop}%
\bibitem [{\citenamefont {Grassini}(2023)}]{grassini2023}%
  \BibitemOpen
  \bibfield  {author} {\bibinfo {author} {\bibfnamefont {S.}~\bibnamefont
  {Grassini}},\ }\bibfield  {title} {\bibinfo {title} {Shaping the future of
  education: exploring the potential and consequences of {AI} and {ChatGPT} in
  educational settings},\ }\href@noop {} {\bibfield  {journal} {\bibinfo
  {journal} {Education Sciences}\ }\textbf {\bibinfo {volume} {13}},\ \bibinfo
  {pages} {692} (\bibinfo {year} {2023})}\BibitemShut {NoStop}%
\bibitem [{\citenamefont
  {Kortemeyer}(2023{\natexlab{c}})}]{kortemeyer24aigrading}%
  \BibitemOpen
  \bibfield  {author} {\bibinfo {author} {\bibfnamefont {G.}~\bibnamefont
  {Kortemeyer}},\ }\bibfield  {title} {\bibinfo {title} {Toward {AI} grading of
  student problem solutions in introductory physics: A feasibility study},\
  }\href {https://doi.org/10.1103/PhysRevPhysEducRes.19.020163} {\bibfield
  {journal} {\bibinfo  {journal} {Phys. Rev. Phys. Educ. Res.}\ }\textbf
  {\bibinfo {volume} {19}},\ \bibinfo {pages} {020163} (\bibinfo {year}
  {2023}{\natexlab{c}})}\BibitemShut {NoStop}%
\bibitem [{\citenamefont {Mori}\ \emph {et~al.}(1992)\citenamefont {Mori},
  \citenamefont {Suen},\ and\ \citenamefont {Yamamoto}}]{mori1992historical}%
  \BibitemOpen
  \bibfield  {author} {\bibinfo {author} {\bibfnamefont {S.}~\bibnamefont
  {Mori}}, \bibinfo {author} {\bibfnamefont {C.~Y.}\ \bibnamefont {Suen}},\
  and\ \bibinfo {author} {\bibfnamefont {K.}~\bibnamefont {Yamamoto}},\
  }\bibfield  {title} {\bibinfo {title} {Historical review of ocr research and
  development},\ }\href@noop {} {\bibfield  {journal} {\bibinfo  {journal}
  {Proceedings of the IEEE}\ }\textbf {\bibinfo {volume} {80}},\ \bibinfo
  {pages} {1029} (\bibinfo {year} {1992})}\BibitemShut {NoStop}%
\bibitem [{\citenamefont {Okamura}\ \emph {et~al.}(1999)\citenamefont
  {Okamura}, \citenamefont {Kanahori}, \citenamefont {Cong}, \citenamefont
  {Fukuda}, \citenamefont {Tamari},\ and\ \citenamefont
  {Suzuki}}]{okamura1999handwriting}%
  \BibitemOpen
  \bibfield  {author} {\bibinfo {author} {\bibfnamefont {H.}~\bibnamefont
  {Okamura}}, \bibinfo {author} {\bibfnamefont {T.}~\bibnamefont {Kanahori}},
  \bibinfo {author} {\bibfnamefont {W.}~\bibnamefont {Cong}}, \bibinfo {author}
  {\bibfnamefont {R.}~\bibnamefont {Fukuda}}, \bibinfo {author} {\bibfnamefont
  {F.}~\bibnamefont {Tamari}},\ and\ \bibinfo {author} {\bibfnamefont
  {M.}~\bibnamefont {Suzuki}},\ }\bibfield  {title} {\bibinfo {title}
  {Handwriting interface for computer algebra systems},\ }in\ \href@noop {}
  {\emph {\bibinfo {booktitle} {Proceedings of the Fourth Asian Technology
  Conference on Mathematics}}}\ (\bibinfo {year} {1999})\ pp.\ \bibinfo {pages}
  {291--300}\BibitemShut {NoStop}%
\bibitem [{\citenamefont {Wang}\ \emph {et~al.}(2021)\citenamefont {Wang},
  \citenamefont {Pan}, \citenamefont {Guo}, \citenamefont {Ji},\ and\
  \citenamefont {Deng}}]{wang2021}%
  \BibitemOpen
  \bibfield  {author} {\bibinfo {author} {\bibfnamefont {H.}~\bibnamefont
  {Wang}}, \bibinfo {author} {\bibfnamefont {C.}~\bibnamefont {Pan}}, \bibinfo
  {author} {\bibfnamefont {X.}~\bibnamefont {Guo}}, \bibinfo {author}
  {\bibfnamefont {C.}~\bibnamefont {Ji}},\ and\ \bibinfo {author}
  {\bibfnamefont {K.}~\bibnamefont {Deng}},\ }\bibfield  {title} {\bibinfo
  {title} {From object detection to text detection and recognition: A brief
  evolution history of optical character recognition},\ }\href@noop {}
  {\bibfield  {journal} {\bibinfo  {journal} {Wiley Interdisciplinary Reviews:
  Computational Statistics}\ }\textbf {\bibinfo {volume} {13}},\ \bibinfo
  {pages} {e1547} (\bibinfo {year} {2021})}\BibitemShut {NoStop}%
\bibitem [{\citenamefont {{Mathpix}}(2024)}]{mathpix}%
  \BibitemOpen
  \bibfield  {author} {\bibinfo {author} {\bibnamefont {{Mathpix}}},\
  }\href@noop {} {\bibinfo {title} {{Mathpix OCR API for STEM}}},\ \bibinfo
  {howpublished} {\url{https://mathpix.com/ocr}} (\bibinfo {year} {accessed
  April 2024})\BibitemShut {NoStop}%
\bibitem [{\citenamefont {Vaswani}\ \emph {et~al.}(2017)\citenamefont
  {Vaswani}, \citenamefont {Shazeer}, \citenamefont {Parmar}, \citenamefont
  {Uszkoreit}, \citenamefont {Jones}, \citenamefont {Gomez}, \citenamefont
  {Kaiser},\ and\ \citenamefont {Polosukhin}}]{vaswani2017attention}%
  \BibitemOpen
  \bibfield  {author} {\bibinfo {author} {\bibfnamefont {A.}~\bibnamefont
  {Vaswani}}, \bibinfo {author} {\bibfnamefont {N.}~\bibnamefont {Shazeer}},
  \bibinfo {author} {\bibfnamefont {N.}~\bibnamefont {Parmar}}, \bibinfo
  {author} {\bibfnamefont {J.}~\bibnamefont {Uszkoreit}}, \bibinfo {author}
  {\bibfnamefont {L.}~\bibnamefont {Jones}}, \bibinfo {author} {\bibfnamefont
  {A.~N.}\ \bibnamefont {Gomez}}, \bibinfo {author} {\bibfnamefont
  {{\L}.}~\bibnamefont {Kaiser}},\ and\ \bibinfo {author} {\bibfnamefont
  {I.}~\bibnamefont {Polosukhin}},\ }\bibfield  {title} {\bibinfo {title}
  {Attention is all you need},\ }\href@noop {} {\bibfield  {journal} {\bibinfo
  {journal} {Advances in neural information processing systems}\ }\textbf
  {\bibinfo {volume} {30}} (\bibinfo {year} {2017})}\BibitemShut {NoStop}%
\bibitem [{\citenamefont {Renze}\ and\ \citenamefont
  {Guven}(2024)}]{renze2024effect}%
  \BibitemOpen
  \bibfield  {author} {\bibinfo {author} {\bibfnamefont {M.}~\bibnamefont
  {Renze}}\ and\ \bibinfo {author} {\bibfnamefont {E.}~\bibnamefont {Guven}},\
  }\bibfield  {title} {\bibinfo {title} {The effect of sampling temperature on
  problem solving in large language models},\ }\href@noop {} {\bibfield
  {journal} {\bibinfo  {journal} {arXiv preprint arXiv:2402.05201}\ } (\bibinfo
  {year} {2024})}\BibitemShut {NoStop}%
\bibitem [{\citenamefont {{Microsoft}}(2024)}]{azure}%
  \BibitemOpen
  \bibfield  {author} {\bibinfo {author} {\bibnamefont {{Microsoft}}},\
  }\href@noop {} {\bibinfo {title} {{Azure AI Services}}},\ \bibinfo
  {howpublished} {\url{https://azure.microsoft.com/en-us/products/ai-services}}
  (\bibinfo {year} {accessed June 2024})\BibitemShut {NoStop}%
\bibitem [{\citenamefont {{R Core Team}}(2021)}]{rproject}%
  \BibitemOpen
  \bibfield  {author} {\bibinfo {author} {\bibnamefont {{R Core Team}}},\
  }\href {https://www.R-project.org/} {\emph {\bibinfo {title} {R: A Language
  and Environment for Statistical Computing}}},\ \bibinfo {organization} {R
  Foundation for Statistical Computing},\ \bibinfo {address} {Vienna, Austria}
  (\bibinfo {year} {2021})\BibitemShut {NoStop}%
\bibitem [{\citenamefont {Fruchterman}\ and\ \citenamefont
  {Reingold}(1991)}]{fruchterman1991}%
  \BibitemOpen
  \bibfield  {author} {\bibinfo {author} {\bibfnamefont {T.~M.}\ \bibnamefont
  {Fruchterman}}\ and\ \bibinfo {author} {\bibfnamefont {E.~M.}\ \bibnamefont
  {Reingold}},\ }\bibfield  {title} {\bibinfo {title} {Graph drawing by
  force-directed placement},\ }\href@noop {} {\bibfield  {journal} {\bibinfo
  {journal} {Software: Practice and experience}\ }\textbf {\bibinfo {volume}
  {21}},\ \bibinfo {pages} {1129} (\bibinfo {year} {1991})}\BibitemShut
  {NoStop}%
\bibitem [{\citenamefont {Epskamp}\ \emph {et~al.}(2012)\citenamefont
  {Epskamp}, \citenamefont {Cramer}, \citenamefont {Waldorp}, \citenamefont
  {Schmittmann},\ and\ \citenamefont {Borsboom}}]{qgraph}%
  \BibitemOpen
  \bibfield  {author} {\bibinfo {author} {\bibfnamefont {S.}~\bibnamefont
  {Epskamp}}, \bibinfo {author} {\bibfnamefont {A.~O.~J.}\ \bibnamefont
  {Cramer}}, \bibinfo {author} {\bibfnamefont {L.~J.}\ \bibnamefont {Waldorp}},
  \bibinfo {author} {\bibfnamefont {V.~D.}\ \bibnamefont {Schmittmann}},\ and\
  \bibinfo {author} {\bibfnamefont {D.}~\bibnamefont {Borsboom}},\ }\bibfield
  {title} {\bibinfo {title} {{qgraph}: Network visualizations of relationships
  in psychometric data},\ }\href {http://www.jstatsoft.org/v48/i04/} {\bibfield
   {journal} {\bibinfo  {journal} {Journal of Statistical Software}\ }\textbf
  {\bibinfo {volume} {48}},\ \bibinfo {pages} {1} (\bibinfo {year}
  {2012})}\BibitemShut {NoStop}%
\bibitem [{\citenamefont {{OpenAI}}(2024{\natexlab{c}})}]{gpt4o}%
  \BibitemOpen
  \bibfield  {author} {\bibinfo {author} {\bibnamefont {{OpenAI}}},\
  }\href@noop {} {\bibinfo {title} {{Hello GPT-4o}}},\ \bibinfo {howpublished}
  {\url{https://openai.com/index/hello-gpt-4o/}} (\bibinfo {year} {accessed
  June 2024}{\natexlab{c}})\BibitemShut {NoStop}%
\end{thebibliography}%

\end{document}